\documentclass[12pt]{article}
\usepackage[mag=1000]{geometry}
\usepackage[fleqn]{amsmath}
\usepackage{amsthm,amssymb,bm,enumerate}
\usepackage{epsf}

\newcommand\bmR{\bm{R}}

\newcommand\bmd{\bm{d}}

\newcommand\bmg{\bm{g}}

\newcommand\bmr{\bm{r}}

\newcommand\bmu{\bm{u}}

\newcommand\bmx{\bm{x}}

\newcommand\bxi{\bm{\xi}}
\newcommand\bOmega{\bm{\Omega}}

\newcommand\bnabla{\bm{\nabla}}

\newcommand\rmc{\mathrm{c}}
\newcommand\rmd{\mathrm{d}}
\newcommand\rme{\mathrm{e}}

\newcommand\rmi{\mathrm{i}}

\newcommand\rmo{\mathrm{o}}
\newcommand\rmp{\mathrm{p}}
\newcommand\rms{\mathrm{s}}
\newcommand\rmt{\mathrm{t}}

\newcommand\imag{\mathrm{Im}}

\newcommand\f{\frac}
\newcommand\p{\partial}

\DeclareMathOperator\sgn{sgn}

\renewcommand\le{\leqslant}
\renewcommand\ge{\geqslant}

\newcounter{questionno}
\begin{document}

\centerline{\textit{\scriptsize To be published in Annual Review of Astronomy \& Astrophysics, Volume 52}}

\bigskip

\centerline{\textbf{\Large Tidal Dissipation in Stars and Giant Planets}}

\bigskip

\centerline{Gordon I.\ Ogilvie}

\medskip

\centerline{\small Department of Applied Mathematics and Theoretical Physics, University of Cambridge}
\centerline{\small Centre for Mathematical Sciences, Wilberforce Road, Cambridge CB3 0WA, UK}
\centerline{\texttt{\small gio10@cam.ac.uk}}

\bigskip

\textbf{Keywords}: celestial mechanics, fluid dynamics, internal
waves, binary stars, extrasolar planets, giant planets: satellites

\bigskip

\textbf{Abstract}: Astrophysical fluid bodies that orbit close to one
another induce tidal distortions and flows that are subject to
dissipative processes.  The spin and orbital motions undergo a coupled
evolution over astronomical timescales, which is relevant for many
types of binary star, short-period extrasolar planetary systems and
the satellites of the giant planets in the solar system.  I review the
principal mechanisms that have been discussed for tidal dissipation in
stars and giant planets in both linear and nonlinear regimes.  I also
compare the expectations based on theoretical models with recent
observational findings.

\bigskip

\textbf{\large 1.\ Introduction}

\medskip

The orbital motion of a gravitating system of extended fluid bodies,
such as stars and giant planets, differs from that of a set of point
masses.  They are non-spherical as a result of both their rotation and
their tidal deformation due to the non-uniform gravitational
attraction of their companions.  In a Newtonian system of two such
bodies, the Keplerian orbital elements evolve in time.  The largest
effects are non-dissipative in character and include orbital
precession.  Of greater interest and subtlety is the irreversible
evolution of the size and shape of the orbit driven by dissipative
processes.

The study of tides has a long and fascinating history (Cartwright
1999; Deparis et al.\ 2013).  It involves an interesting combination
of celestial mechanics, which is non-trivial but can be explored
relatively easily, and fluid dynamics, which involves some deeper
physical issues that will be touched on this article.  Indeed, this
review is devoted to stars and giant planets, which are wholly or
predominantly fluid, rather than terrestrial bodies, which are
predominantly solid, although deformable.  In other words, our
interest is in fluid dynamics rather than solid mechanics, in which
the dissipative properties have a different nature.  However, there is
not a clean separation between these areas.  Neutron stars have solid
crusts, giant planets may contain solid cores, and terrestrial bodies
may contain oceans and atmospheres with important tidal phenomena.  In
the case of the Earth, a fluid layer that is only $0.023\%$ of the
planet's mass dominates the tidal dissipation.

Our focus is also on purely gravitational tides, which can be regarded
as a subset of interactions between neighbouring astrophysical bodies.
Thermal and magnetic tides are also possible, in which one body
affects the other in a way that is not spherically symmetric and may
depend periodically on time owing to the spin and orbital motion.  As
for gravitational tides, both wavelike and non-wavelike disturbances
may be generated.

\begin{figure}
\centerline{\epsfysize10cm\epsfbox{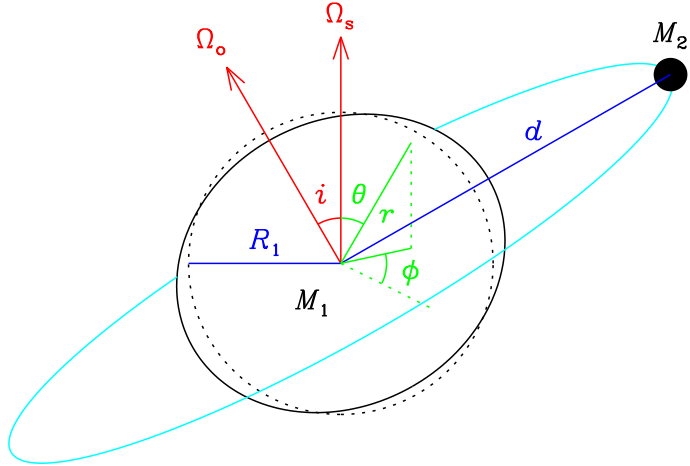}}
\vskip-2cm
\caption{Geometry of tidal interaction.  When considering the tide
  raised in body~1 by body~2, it is convenient to use spherical polar
  coordinates $(r,\theta,\phi)$ centred on body~1 and aligned with its
  rotation axis.  $\bOmega_\rms$ and $\bOmega_\rmo$ are the spin and
  orbital angular velocity vectors.  The figure represents the tidal
  bulge but not the rotational bulge.}
\label{f:geometry}
\end{figure}

A useful dimensionless parameter that can be taken as a simple
estimate of the tidal deformation of one body by another is the tidal
amplitude parameter $\epsilon=(M_2/M_1)(R_1/d)^3$, where $M_1$ and
$M_2$ are the masses of the two bodies, $R_1$ is the (mean) radius of
body 1 and $d$ is the orbital separation (Fig.~\ref{f:geometry}).
This is a measure of the ratio, at the surface of body~1, of the tidal
gravity due to body~2, $GM_2R_1/d^3$, to the gravity of body~1 itself,
$GM_1/R_1^2$; it is also an estimate of the ratio of the height of the
tide compared to the radius of body~1.

Tidal interactions are therefore strongly dependent on the separation
of the bodies in comparison with their sizes.  The most familiar example
(although beyond the scope of this article) is of course the
Earth--Moon system, in which the Sun also plays a significant role.
Tidal dissipation in the Earth leads to a lengthening of both the day
and the month as angular momentum is transferred from the planetary
spin to the orbit.  Tidal evolution, including the effects of tidal
dissipation in giant planets, has been similarly influential for many
of the regular satellites of the other planets in the solar system
(Peale 1999).  Tidal interactions have many applications to binary
stars, including pre-main-sequence, main-sequence, giant and
degenerate stars, whenever the ratio of orbital separation to stellar
radius is sufficiently small.  Much recent work on tides has been
stimulated by the discovery since 1995 of numerous extrasolar planets
that orbit very close to their host stars, and also by the prospect of
detecting gravitational radiation from merging double-degenerate
binary stars, where tidal effects may alter the predicted wave signals
(e.g.\ Kochanek 1992; Bildsten \& Cutler 1992).

In this review we concentrate on situations in which the tidal
amplitude parameter $\epsilon\ll1$, although important internal
nonlinearities may still be present, as discussed in Section~4. (There
are of course more extreme situations in which the body is tidally
disrupted, resulting in a significant loss of mass; this includes
Roche-lobe overflow in binary stars with circular orbits and the tidal
disruption of stars on eccentric orbits around the black hole at the
centre of a galaxy, or of planets scattered close to their host stars,
e.g.\ Guillochon et al.\ 2011).  For the giant planets of the solar
system, the largest values of $\epsilon$ are approximately
$2\times10^{-7}$ (Jupiter, due to Io), $3\times10^{-8}$ (Saturn, due
to Titan), $4\times10^{-8}$ (Uranus, due to Ariel) and
$8\times10^{-8}$ (Neptune, due to Triton).  Among exoplanetary systems
confirmed so far, the largest values are approximately
$2\times10^{-4}$ for the tide in a star (WASP-18; Hellier et al.\
2009) and $6\times10^{-2}$ for the tide in a planet (WASP-19~b; Hebb
et al.\ 2010).  The tides in giant planets forced by small satellites
and by nearby host stars may be in very different regimes, which
motivates a study of linear and nonlinear tides.

The dominant tidal interaction is usually between the gravitational
quadrupole moment of one body and the monopole moment of the other.
The second body can then be thought of as a point mass, while the
first is endowed with an ellipsoidal bulge (Fig.~\ref{f:geometry}).
The monopole field of body~2 varies over the volume occupied by
body~1, creating a tidal field that is aligned with the axis joining
the centres of the bodies.  This tidal field deforms body~1,
generating a quadrupole moment.  In the simplest case a spheroidal
bulge arises, aligned with the axis, which gives rise to an attractive
net force in that radial direction.

The process of deformation involves fluid dynamics, as discussed in
detail later in this article, and is generally accompanied by
dissipation.  As a result, the quadrupole moment of body~1 is not
instantaneously related to the tidal field and is not generally
aligned with the radial direction.  This effect is often modelled
crudely by allowing for a certain time lag in the response.  The
dissipative part of the net force between the bodies has two aspects.
One is that the radial force is retarded and therefore resists the
oscillatory radial motion associated with any eccentricity of the
orbit.  The other is that an angular force arises that resists the
relative angular motion associated with any inequality of the spin and
orbital angular velocity vectors.  This force also gives rise to a
tidal torque that couples the spin and orbit, allowing an exchange of
angular momentum.  Both forces are resistive in nature and are
accompanied by dissipation of energy.

The responses of spin and orbit to changes in angular momentum are
different.  While spin angular momentum increases with angular
velocity, orbital angular momentum decreases, which can lead to
paradoxical behaviour.  The tidal torque seeks to equalize the the
spin and orbital angular velocity vectors through an exchange of
angular momentum and the dissipation of energy.  In many cases the
spin moment of inertia is much smaller than the orbital moment of
inertia, and the first effect of tidal dissipation is a tendency
towards equalization of the spin and orbital angular velocity vectors,
involving only a small change in the orbit.  In the case of a circular
orbit this means both synchronization of the spin with the orbit and
alignment of the spin axis with the orbit normal, with the final
angular velocity vector being determined by the total angular
momentum.  If the orbit is eccentric, however, then the orbital
angular velocity is not constant and there is a tendency towards
`pseudo-synchronization', i.e.\ spin at a rate close to, but slightly
less than, the instantaneous orbital angular velocity at the
pericentre, where the tidal interaction is strongest (Hut 1981).

Both the radial force and the tidal torque contribute to the evolution
of eccentricity, which can either increase or decrease as a result.
Similarly, the obliquity, which is the angle between the spin and
orbital angular momentum vectors, can either increase or decrease.
Circularization and spin-orbit alignment may be regarded as the normal
outcomes of tidal dissipation; in order to increase $e$ or $i$, the
energy source associated with an asynchronous spin is needed.  This
possibility is discussed in more detail in Section~2.3 below.

In most binary stars the theoretical endpoint of tidal evolution, if
other processes such as magnetic braking, gravitational radiation and
stellar evolution are ignored, is a tidal equilibrium, or double
synchronous state, in which both stars are in aligned, synchronous
rotation with a circular orbit, as occurs with Pluto and Charon.
There is then only a static tidal deformation and no dissipation.
(However, internal processes such as convection and meridional
circulation may induce a differential rotation in the stars, which
would imply ongoing tidal dissipation.)  The double synchronous state
has not yet been verified observationally for a binary star.  Two
candidates can be found in Table~1 of Meibom et al.\ (2006), although
in each binary only one spin period and neither obliquity is measured.
Albrecht et al.\ (2007) give the first observations of both stellar
obliquities (consistent with zero) in a main-sequence binary, but the
orbit is eccentric.

In systems of extreme mass ratio, however, such as the satellite
systems of the giant planets and many short-period extrasolar
planetary systems, this endpoint cannot be reached.  Tidal equilibria
may be characterized by their angular velocity $\Omega$.  The total
angular momentum $L(\Omega)$ of a tidal equilibrium is the sum of the
spin and orbital contributions, $L_\mathrm{s}(\Omega)$ and
$L_\mathrm{o}(\Omega)$, which are increasing and decreasing functions,
respectively, such that $L(\Omega)$ has a minimum, critical value
$L_\mathrm{c}$ at $\Omega=\Omega_\mathrm{c}$.  If $L<L_\mathrm{c}$
then no tidal equilibrium is accessible.  If $L>L_\mathrm{c}$ then two
such states exist.  The less compact one with
$\Omega<\Omega_\mathrm{c}$ has $L_\mathrm{o}>3L_\mathrm{s}$ and is
stable, being a minimum of the energy subject to the angular momentum
being conserved (Counselman 1973; Hut 1980).  The more compact one
with $\Omega>\Omega_\mathrm{c}$ has $L_\mathrm{o}<3L_\mathrm{s}$ and
is unstable; it may in any case be inaccessible if the orbit would be
too small relative to the sizes of the bodies.  The expressions for
$L_\rmc$ and $\Omega_\rmc$ are
\begin{equation}
  L_\rmc=4I\Omega_\rmc,\qquad
  \Omega_\rmc=(GM)^{1/2}\left(\f{\mu}{3I}\right)^{3/4},
\label{lcoc}
\end{equation}
where $M=M_1+M_2$ is the total mass, $\mu=M_1M_2/M$ is the reduced
mass, and $I=I_1+I_2$ is the sum of the spin moments of inertia
(assumed constant).

Many of the short-period extrasolar planetary systems for which tidal
interactions are most important (having orbital periods less than
about five days) have $L<L_\mathrm{c}$ and therefore no tidal
equilibrium; the orbit shrinks until the planet is destroyed. A few
short-period systems with very massive planets are estimated to have
$L>L_\mathrm{c}$ (Levrard et al.\ 2009; Matsumura et al.\ 2010).  In
two of these, CoRoT-3~b and $\tau$~Boo~b, the stellar spin appears to
be approximately synchronized with the orbit (Husnoo et al.\ 2012) and
these systems may already be in, or close to, a stable tidal
equilibrium.  In the case of $\tau$~Boo~b (not mentioned by Levrard et
al.\ or Matsumura et al.) it is likely that $L>L_\mathrm{c}$ by only a
narrow margin; the same is true of KELT-1 (Siverd et al.\ 2012) and
CoRoT-15 (Bouchy et al.\ 2011), whose companions would usually be
classified as brown dwarfs rather than planets.  Some other
short-period systems with $L>L_\mathrm{c}$ could be evolving towards a
stable equilibrium, although magnetic braking may eventually reduce
$L$ below $L_\mathrm{c}$ and destroy the equilibrium.  Many
longer-period systems with massive planets also have $L>L_\mathrm{c}$
but are not expected to undergo significant tidal evolution.

All satellite systems of solar-system planets have
$L>L_\mathrm{c}$. For the giant planets, $L\gg L_\mathrm{c}$ and
$L_\mathrm{o}\ll3L_\mathrm{s}$; if a synchronously spinning satellite
were placed in a circular orbit near the corotation radius
(synchronous orbit) of a giant planet, it would migrate away from that
radius and the planetary spin would not be greatly affected.

Tidal dissipation generates heat in astrophysical bodies, which in
some cases may be important for their structure and evolution.  Apart
from the well known applications to the volcanic activity of Jupiter's
moon Io and Saturn's moon Enceladus, this effect has been investigated
mainly for short-period extrasolar planets undergoing tidal
circularization, in an attempt to explain the unexpectedly large radii
of many transiting planets (e.g.\ Bodenheimer et al.\ 2001; Ibgui \&
Burrows 2009).  In fact, the orbital energy that must be dissipated to
circularize a short-period planet can easily exceed the planet's
internal binding energy, which suggests that rapid circularization, if
possible, either makes the planet very bright or threatens to destroy
it.  Gu et al.\ (2003) identified a tidal inflation instability that
may lead to the disruption of gas giants.

\bigskip

\textbf{\large 2.\ Tidal dynamics}

\medskip

\textbf{2.1.\ Tidal potential}

Consider two bodies orbiting about their mutual centre of mass.  Let
their masses be $M_1$ and $M_2$, and their centres of mass $\bmR_1(t)$
and $\bmR_2(t)$.  Suppose that body~2 is a point mass, or can be
treated as such for the purposes of determining the motion and tidal
deformation of body~2.  When its gravitational potential
$-GM_2/|\bmr-\bmR_2|$ is expanded in a Taylor series about
$\bmr=\bmR_1$, we obtain
\begin{equation}
  -\f{GM_2}{|\bmd|}\left[1+\f{\bmd\cdot\bmx}{|\bmd|^2}+\f{3(\bmd\cdot\bmx)^2-|\bmd|^2|\bmx|^2}{2|\bmd|^4}+\cdots\right]=-\f{GM_2}{|\bmd|}\sum_{l=0}^\infty\f{|\bmx|^l}{|\bmd|^l}P_l\left(\f{\bmd\cdot\bmx}{|\bmd||\bmx|}\right),
\end{equation}
where $\bmd=\bmR_2-\bmR_1$ is the orbital separation,
$\bmx=\bmr-\bmR_1$ is the position vector with respect to the centre
of body~1, and $P_l$ is the Legendre polynomial of degree~$l$.  The
first term in this series is a uniform potential and has no effect,
while the second term gives rise to a uniform acceleration, which
causes the basic Keplerian orbital motion of body~1.  The remaining
part of the expansion defines the tidal potential $\Psi$, which gives
rise to a non-uniform acceleration that deforms body~1.  It can also
be represented using solid spherical harmonic functions of the second
degree and higher, i.e.\ $r^lY_l^m(\theta,\phi)$ with $l\ge2$ and
$|m|\le l$, where $(r,\theta,\phi)$ are spherical polar coordinates
with their origin at the centre of body~1.

If body~1 is rotating, then for the purposes of computing the tidal
deformation it is helpful to choose the axis of the coordinate system
to coincide with the rotation axis.  Let the orbit have semi-major axis
$a$, eccentricity $e$ and inclination $i$ with respect to the
equatorial plane of body~1.  In celestial mechanics $i$ is known as
the obliquity of body~1 (with respect to the orbit of bodies~1 and~2).
The periodic variation of the separation vector due to the basic
Keplerian orbital motion can be expressed through Fourier series,
leading to an expansion of the complete tidal potential in a
non-rotating frame in the form (cf.\ Kaula 1961; Polfliet \& Smeyers
1990)
\begin{equation}
  \Psi=\mathrm{Re}\sum_{l=2}^\infty\sum_{m=0}^l\sum_{n=-\infty}^\infty\f{GM_2}{a}A_{l,m,n}(e,i)\left(\f{r}{a}\right)^lY_l^m(\theta,\phi)\,\rme^{-\rmi n\Omega_\rmo t}.
\end{equation}
Here $\Omega_\rmo=(GM/a^3)^{1/2}$ is the mean orbital angular
velocity, which we will refer to as the `orbital frequency' (and is
known as the mean motion in celestial mechanics).  The dimensionless
complex coefficients $A_{l,m,n}$ depend in a complicated way on $e$
and $i$.  The integers $l$ and $m$ are the degree and order of the
spherical harmonic; $m$ is also referred to as the azimuthal
wavenumber.  The integer $n$ labels temporal harmonics of the orbital
motion.

In most applications the bodies are sufficiently well separated that
the quadrupolar components ($l=2$) are strongly dominant.  In the
special case of a circular, coplanar orbit ($e=i=0$) the only terms
present have $n=m$, and $l-m$ must be even.  In the case of a
circular, inclined orbit ($e=0$) $n$ is restricted to the range
$[-l,l]$, and $l-n$ must be even (e.g.\ Ogilvie 2013).  For the
complete representation of an eccentric orbit, all values of $n$ are
required.  However, if terms smaller than $O(e^p)$ can be neglected,
then the largest value of $|n|$ that need be considered is $l+p$.
Table~\ref{t:components} gives the amplitudes, but not the phases, of
the quadrupolar components of the tidal potential, correct to first
order in $e$ and $i$.

\begin{table}
\caption{Quadrupolar components of the tidal potential, correct to first order in eccentricity and obliquity.}
\begin{center}
\begin{tabular}{cccccc}
$l$&$m$&$n$&$|A|$&description\\
\\
$2$&$0$&$0$&$\sqrt\f{\pi}{5}$&static tide\\
$2$&$2$&$2$&$\sqrt\f{6\pi}{5}$&asynchronous tide\\
$2$&$0$&$1$&$3e\sqrt\f{\pi}{5}$&eccentricity tides\\
$2$&$2$&$1$&$\f{1}{2}e\sqrt\f{6\pi}{5}$&\\
$2$&$2$&$3$&$\f{7}{2}e\sqrt\f{6\pi}{5}$&\\
$2$&$1$&$0$&$i\sqrt\f{6\pi}{5}$&obliquity tides\\
$2$&$1$&$2$&$i\sqrt\f{6\pi}{5}$&\\
\end{tabular}
\end{center}
\label{t:components}
\end{table}

Since $Y_l^m(\theta,\phi)\propto\rme^{\rmi m\phi}$, the phase of each
tidal component is $\arg A_{l,m,n}+m\phi-n\Omega_\rmo t$.  When
$m\ne0$, the phase rotates with angular velocity $n\Omega_\rmo/m$.
The angular frequency of each component measured in a non-rotating
frame is $\omega=n\Omega_\rmo$, which may be called the `tidal
frequency in the inertial frame'.  Of greater importance for the
physical response of the fluid is the angular frequency measured in a
frame that rotates with the spin angular velocity $\Omega_\rms$ (`spin
frequency') of body~1, $\hat\omega=n\Omega_\rmo-m\Omega_\rms$, which
may be called the `tidal frequency in the fluid frame'.  When $m\ne0$,
the difference between $\omega$ and $\hat\omega$ is due to an angular
Doppler shift.  (In general, body~1 may rotate differentially, in
which case $\hat\omega$ depends on position.)

\begin{figure}
\centerline{\epsfysize10cm\epsfbox{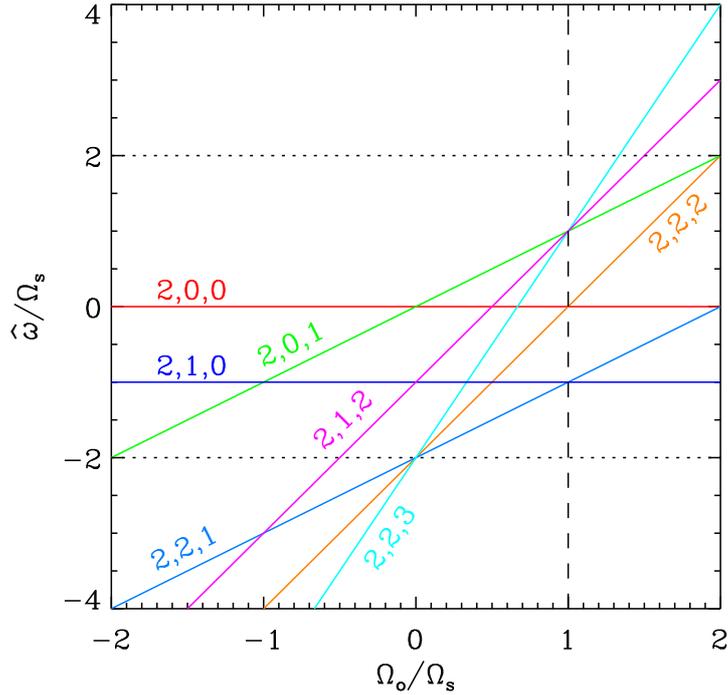}}
\caption{Tidal forcing frequencies for the seven lowest-order
  components listed in Table~\ref{t:components}, labelled by $l,m,n$.
  The vertical axis is the ratio of the tidal frequency in the fluid
  frame to the spin frequency, while the horizontal axis is the ratio
  of the orbital frequency to the spin frequency.  Each line has slope
  $n$ and $y$-intercept $-m$.  The dotted horizontal lines indicate
  the frequency range of inertial waves in a uniformly rotating body
  (see Section~3.5 below), while the dashed vertical line indicates
  the synchronous state.}
\label{f:frequencies}
\end{figure}

The tidal frequencies in the fluid frame are therefore integer linear
combinations of the spin and orbital frequencies; for the seven
components listed in Table~\ref{t:components}, these are plotted in
Fig.~\ref{f:frequencies}.  As is discussed in Section~3, the Coriolis
force plays a dominant role in the wavelike part of the tidal response
when $|\hat\omega/\Omega_\rms|<2$ (and may still play an important
role outside this interval), and it can be seen that all seven
components typically have frequencies in this range, unless the body
is far from the synchronous state.

\begin{figure}
\centerline{\epsfysize10cm\epsfbox{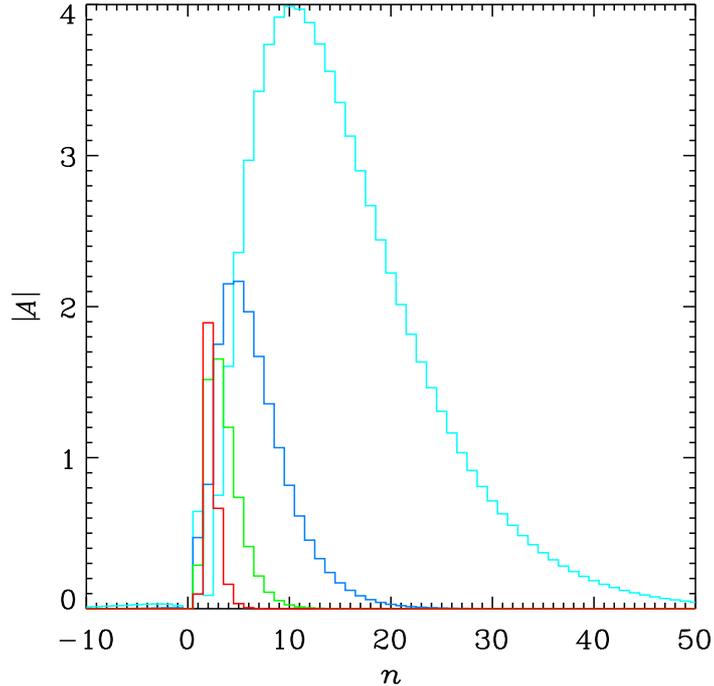}}
\caption{Amplitude $|A_{2,2,n}|$ of the $l=m=2$ component of the tidal
  potential for obliquity $i=0$ and eccentricities $e=0.1$, $0.3$,
  $0.5$ and $0.7$ (red, green, blue and cyan lines)}
\label{f:eccentric}
\end{figure}

As the eccentricity is increased from $0$ towards $1$, the
time-dependence of the tidal forcing changes from a sinusoidal
variation to one that is strongly peaked at the pericentre of the
orbit.  This is a result of the sensitivity of the tidal force to the
orbital separation.  The above expansion is still valid for large
eccentricities, but a broad spectrum of frequencies is obtained
(Fig.~\ref{f:eccentric}).  In a highly eccentric orbit, the tidal
interaction has an impulsive character, consisting of a series of
tidal `encounters', each of which might be approximated as a parabolic
orbit.

\newpage

\textbf{2.2.\ Tidal torque, power and dissipation}

If the tidal amplitude parameter $\epsilon$ is sufficiently small, the
tidal response of a body can be determined using linear theory, as
discussed in Section~3 below.  In this approach, the tidal disturbance
is treated as a small perturbation of a basic state, which is usually
assumed to be steady and axisymmetric.  To each component of the tidal
potential there is an independent and linearly related response, which
can be quantified using a response function.  If a tidal potential
component
$\mathrm{Re}[\mathcal{A}(r/R)^lY_l^m(\theta,\phi)\,\rme^{-\rmi\omega
  t}]$ is applied to a body of nominal (e.g.\ mean or equatorial)
radius $R$, and the resulting deformation of the body generates an
external gravitational potential perturbation
$\mathrm{Re}[\mathcal{B}(R/r)^{l+1}Y_l^m(\theta,\phi)\,\rme^{-\rmi\omega
  t}]$ (possibly in addition to other components that are orthogonal
to the applied one), then the complex dimensionless ratio
$k_l^m(\omega)=\mathcal{B}/\mathcal{A}$ defines the \textit{potential
  Love number} (named after A.~E.~H.~Love).  This is the most useful
response function, because it is only through gravity that the tidal
communication between the two bodies occurs.  Of particular interest is
the imaginary part $\mathrm{Im}[k_l^m(\omega)]$, which quantifies the
part of the response that is out of phase with the tidal forcing and
is associated with transfers of energy and angular momentum.

Let the tidal power $P$ and the tidal torque $T$ be defined as the
rates of transfer of energy and of the axial component of angular
momentum from the orbit to the body (measured in the inertial frame,
and averaged in time in the case of $m=0$).  It can be shown (e.g.\
Ogilvie 2013) that $P=\omega\mathcal{T}$ and $T=m\mathcal{T}$, where
\begin{equation}
  \mathcal{T}=\f{(2l+1)R|\mathcal{A}|^2}{8\pi G}\,\mathrm{Im}[k_l^m(\omega)].
\end{equation}
In a frame that rotates with angular velocity $\Omega$, the torque is
again $T$ but the power is $P-T\Omega$.  When $m\ne0$, the fact that
$P=(\omega/m)T$ means that the power vanishes in a rotating frame in
which the tidal potential is stationary.

If the body rotates uniformly with angular velocity $\Omega_\rms$,
then the (frame-independent) tidal dissipation rate is
$D=P-T\Omega_\rms=\hat\omega\mathcal{T}$.  Although the torque, power
and dissipation rate are all quantities of second order in $\epsilon$,
they can be calculated within a linear analysis and are linearly
proportional to the imaginary part of the response function.  The
difference between the dissipation rate and the power is the rate of
change of the spin energy of the body, which vanishes in the fluid
frame.  Since $D$ is a positive-definite quantity, we have the
important result that $\mathrm{Im}[k_l^m(\omega)]$ has the same sign
as $\hat\omega$.  [A possible exception is that convection could in
principle give energy to the tidal disturbance rather than dissipating
it (Ogilvie \& Lesur 2012).]  $\mathrm{Im}[k_l^m(\omega)]$ should also
vanish when $\hat\omega=0$, because then the deformation is static in
the fluid frame and should be in phase with the tidal forcing.

If the body rotates non-uniformly, then there is no simple global
relation between $D$, $P$ and $T$.  The dissipation rate cannot then
be deduced from $\mathrm{Im}[k_l^m(\omega)]$ and ceases to be of such
direct importance, except of course for tidal heating.

Several other parametrizations of the tidal response are in common
use.  Typically $\mathrm{Re}[k_l^m(\omega)]$ is a quantity of order
unity, only weakly dependent on $m$ and $\omega$, and can be well
approximated by its hydrostatic value.  The evaluation of the
hydrostatic Love number $k_l$ (which is real and does not depend on
$m$) for a spherical fluid body is a classical problem involving the
equations of Clairaut and Radau (e.g.\ Kopal 1978); for a homogeneous
body, $k_l=k_l^\mathrm{hom}=3/[2(l-1)]$.  [The apsidal motion
constants long used in the theory of binary stars (e.g.\ Kopal 1978)
are denoted by the same symbol $k_l$ and are equivalent except that
they are smaller by a factor of $2$.]  It is then common to write
$|\mathrm{Im}[k_l^m(\omega)]|$ as $k_l/Q$, where $Q$ is the
\textit{tidal quality factor}.  It is also common to write $k_l/Q$ as
$k_l^\mathrm{hom}/Q'$, where $Q'$ is a \textit{modified tidal quality
  factor}; this has the advantage of combining $Q$ with $k_l$, which
may not be known accurately as it depends on the interior structure,
whereas $k_l^\mathrm{hom}$ has a simple analytical expression.  [For
giant planets, it is estimated that $0.1<k_2<0.4$ (Gavrilov \& Zharkov
1977; Kramm et al.\ 2012), although smaller values are possible for
very centrally condensed models.  For the Sun, $k_2\approx0.0351$,
making it even more important to distinguish clearly between $Q$ and
$Q'$.]  The \textit{phase lag} in the tidal response is $1/Q$ and the
\textit{time lag} is $\tau=1/(Q|\hat\omega|)$.  To summarize, in the
case of quadrupolar tides ($l=2$) the various parametrizations of
tidal dissipation are related by
\begin{equation}
  \mathrm{Im}[k_2^m(\omega)]=\sigma\f{3}{2Q'}=\sigma\f{k_2}{Q}=k_2\tau\hat\omega,
\end{equation}
where $\sigma=\sgn\hat\omega=\pm1$.

The expressions `quality factor', `phase lag' and `time lag' should
not be taken literally.  The above relations are based on a simplified
conceptual model in which the tidal response resembles the hydrostatic
one, but is slightly retarded in phase and unaffected in amplitude.
However, if the tidal forcing resonates with a weakly damped normal
mode of oscillation, especially if it is the fundamental mode as
discussed in Section~3.1, it is possible in principle for
$|\mathrm{Im}[k_l^m(\omega)]|$ to attain values much larger than unity
because the amplitude of the tidal response greatly exceeds the
hydrostatic one.  No amount of phase lag could describe such a
response.  However, we can continue to use the above formal relations
if we allow $Q$ and $Q'$ to be less than unity in such a situation
(which is, however, unlikely to arise in practice).

Potential Love numbers can still be used to describe the tidal
deformation of a body in a nonlinear regime, but then there can be
interference between different components of the tide and the Love
numbers are no longer linear response functions of the body but depend
on the amplitudes and relative phases of all applied tidal components.

It is particularly important to note that the tidal quality factor $Q$
is not a fundamental material property of a body, but is a
dimensionless parametrization of a response function, which is partly
why we favour the notation $\imag[k_l^m(\omega)]$.  As well as
depending on $l$, $m$ and $\omega$, $\imag(k)$ depends on the
internal structure and angular velocity of the body.  In his first
published paper, Goldreich (1963) issues words of caution about this
parametrization: ``$Q$ will in general vary with the frequency and
amplitude of the tide and the size of the sphere in addition to its
composition\dots In our discussions we shall use the language of
linear tidal theory, but we must keep in mind that our numbers are
really only parametric fits to a non-linear problem.''

\medskip

\textbf{2.3.\ Rate and direction of tidal evolution}

To lowest order in $e$ and $i$, the rates of changes of the orbital
semi-major axis $a$, the spin angular velocity $\Omega_\rms$ of
body~1, the orbital eccentricity $e$ and the obliquity (spin--orbit
misalignment) $i$ of body~1 due to tidal dissipation in body~1 are
given by
\begin{equation}
  \f{1}{a}\f{\rmd a}{\rmd t}=-3\kappa_{2,2,2}\f{M_2}{M_1}\left(\f{R_1}{a}\right)^5\Omega_\rmo,
\label{adot}
\end{equation}
\begin{equation}
  \f{1}{\Omega_\rms}\f{\rmd\Omega_\rms}{\rmd t}=\f{3}{2}\kappa_{2,2,2}\f{L_\rmo}{L_\rms}\f{M_2}{M_1}\left(\f{R_1}{a}\right)^5\Omega_\rmo,
\label{sdot}
\end{equation}
\begin{equation}
  \f{1}{e}\f{\rmd e}{\rmd t}=\f{3}{16}(4\kappa_{2,2,2}-6\kappa_{2,0,1}+\kappa_{2,2,1}-49\kappa_{2,2,3})\f{M_2}{M_1}\left(\f{R_1}{a}\right)^5\Omega_\rmo,
\label{edot}                                                                   
\end{equation}
\begin{equation}
  \f{1}{i}\f{\rmd i}{\rmd t}=\f{3}{4}\left[\kappa_{2,2,2}\left(1-\f{L_\rmo}{L_\rms}\right)+(\kappa_{2,1,0}-\kappa_{2,1,2})\left(1+\f{L_\rmo}{L_\rms}\right)\right]\f{M_2}{M_1}\left(\f{R_1}{a}\right)^5\Omega_\rmo,
\label{idot}
\end{equation}
where we introduce the notation
\begin{equation}
  \kappa_{l,m,n}=\mathrm{Im}[k_l^m(n\Omega_\rmo)]
\end{equation}
for the imaginary part of the Love number corresponding to the tidal
component labelled by $l,m,n$, and
\begin{equation}
  \f{L_\rmo}{L_\rms}=\f{\mu[GMa(1-e^2)]^{1/2}}{I_1\Omega_\rms}=\f{GM_1M_2(1-e^2)^{1/2}}{I_1\Omega_\rms\Omega_\rmo a}
\end{equation}
is the ratio of the orbital angular momentum to the spin angular
momentum of body~1 (if it has a fixed moment of inertia $I_1$).  Note
that these formulae are valid only for sufficiently small values of
$e$ and $i$.  Our notation assumes that $\Omega_\rmo$ and $L_\rmo$ are
positive by definition, as are $a$, $e$ and $i$, whereas
$L_\rms=I_1\Omega_\rms$ can be either positive or negative.  The case
$\Omega_\rms<0$ allows us to consider situations in which spin and
orbit are (nearly) counteraligned.  Results of this type can be found
in the classic work of Darwin (1880), although expressed in a
different notation; they are derived by considering the effects of the
tidal torque and power on the spin and orbit.  The effects of tidal
dissipation in body~2 on the orbital semi-major axis and eccentricity,
and on \textit{its} spin angular velocity and obliquity, are simply
obtained by interchanging the roles of the two bodies in this
calculation.

Without considering yet the details of the tidal responses of
astrophysical bodies, but assuming that $|\kappa|\lesssim1$ (usually
$\ll1$) for each component, a number of observations can be made.
First, tidal evolution occurs on a timescale that is much longer than
the orbital timescale, if $\epsilon\ll1$ as we have assumed.  Second,
tidal dissipation in the two bodies may be of comparable importance;
the ratio of the factor $(M_2/M_1)(R_1/a)^5$ to the corresponding
factor $(M_1/M_2)(R_2/a)^5$ when the roles are reversed can be
expressed in terms of the mean densities of the bodies as
$(\bar\rho_2/\bar\rho_1)^2(R_2/R_1)$, which suggests that dissipation
in the smaller body can dominate the evolution of $e$ (Goldreich
1963), and also of $a$ before body~2 becomes synchronized.  (Among
exoplanetary systems, this ratio has values both less than and greater
than unity.)  Third, in systems for which $L_\rms\ll L_\rmo$, spin
evolution proceeds much more rapidly than orbital evolution.

\begin{figure}
\centerline{\epsfysize10cm\epsfbox{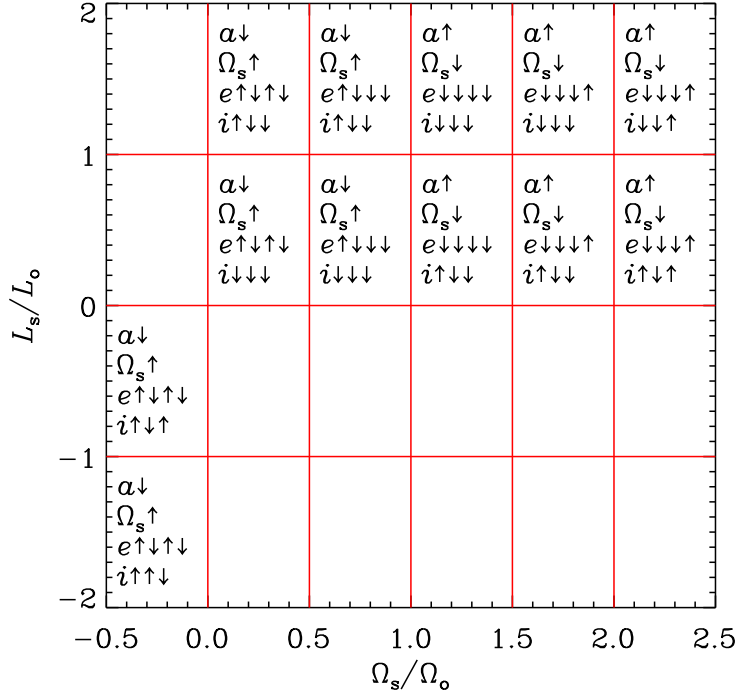}}
\caption{Signs of the contributions of the lowest-order tidal
  components to the rates of change of $a$, $\Omega_\rms$, $e$ and
  $i$.  The horizontal axis is the ratio of the spin frequency to the
  orbital frequency, while the vertical axis is the ratio of the spin
  angular momentum to the orbital angular momentum.  Only the first
  and third quadrants are physically accessible.  Upward and downward
  arrows indicate positive and negative rates of change, respectively.
  In the cases of $e$ and $i$ the contributions are shown in the same
  order as that given in equations~(\ref{edot}) and~(\ref{idot}).  No
  further sign changes occur beyond the region plotted here.}
\label{f:signs}
\end{figure}

If body~1 rotates uniformly with angular velocity $\Omega_\rms$ and
the tidal dissipation rate is positive, then, as noted above,
$\kappa_{l,m,n}$ must have the same sign as the sign of the tidal
frequency in the fluid frame, $\hat\omega=n\Omega_\rmo-m\Omega_\rms$,
and must vanish when $\hat\omega=0$.  Note that $\kappa_{2,2,2}$
vanishes in the synchronous case $\Omega_\rms=\Omega_\rmo$.  Even in
the lowest-order analysis presented here, numerous possibilities arise
for the signs of the above rates of change.  The signs of the
contributions of the lowest-order tidal components to the rates of
change of $a$, $\Omega_\rms$, $e$ and $i$ are summarized in
Fig.~\ref{f:signs}.  In most cases there are competing contributions
to $\rmd e/\rmd t$ and $\rmd i/\rmd t$ and the net result is not
obvious (Jeffreys 1961; Goldreich 1963).  The fourth contribution to
$\rmd e/\rmd t$ comes with the largest weighting factor, but this
might be offset by a weaker tidal response at the corresponding
frequency.

In the unlikely event that all components have the same time lag
$\tau$, which is the case investigated in detail by Darwin (1880),
Alexander (1973), Hut (1981) and many others, we have
$\kappa_{l,m,n}=k_l\tau(n\Omega_\rmo-m\Omega_\rms)$ and obtain the
simpler results
\begin{equation}                                                                 \f{1}{a}\f{\rmd a}{\rmd t}=-6k_2\tau(\Omega_\rmo-\Omega_\rms)\f{M_2}{M_1}\left(\f{R_1}{a}\right)^5\Omega_\rmo,
\label{adot_tau}
\end{equation}
\begin{equation}                                                                 \f{1}{\Omega_\rms}\f{\rmd\Omega_\rms}{\rmd t}=3k_2\tau(\Omega_\rmo-\Omega_\rms)\f{L_\rmo}{L_\rms}\f{M_2}{M_1}\left(\f{R_1}{a}\right)^5\Omega_\rmo,
\label{sdot_tau}
\end{equation}
\begin{equation}
  \f{1}{e}\f{\rmd e}{\rmd t}=-\f{3}{2}k_2\tau(18\Omega_\rmo-11\Omega_\rms)\f{M_2}{M_1}\left(\f{R_1}{a}\right)^5\Omega_\rmo,
\label{edot_tau}
\end{equation}
\begin{equation}
  \f{1}{i}\f{\rmd i}{\rmd t}=-\f{3}{2}k_2\tau\left[\Omega_\rms+(2\Omega_\rmo-\Omega_\rms)\f{L_\rmo}{L_\rms}\right]\f{M_2}{M_1}\left(\f{R_1}{a}\right)^5\Omega_\rmo,                           
\label{idot_tau}
\end{equation}
which would imply that the eccentricity is excited for
$\Omega_\rms/\Omega_\rmo>18/11$ (Darwin 1880), while the obliquity is
excited for $\Omega_\rms/\Omega_\rmo>2L_\rmo/(L_\rmo-L_\rms)$, if
$0<L_\rms<L_\rmo$ (Hut 1981), or always, if $L_\rms<0$ (i.e.\
$\Omega_\rms<0$).  The extension of these equations to arbitrary $e$
and $i$ under the assumption of a common time lag can be found in
Leconte et al.\ (2010) and Matsumura et al.\ (2010), and was first
derived by Alexander (1973).

\medskip

\textbf{\large 3.\ Linear tides}

\medskip

\textbf{3.1.\ Linear response of a star or giant planet to harmonic
  forcing}

Let us consider in more detail the linear response of an astrophysical
fluid body to periodic forcing by a single tidal component.  A simple
explicit example is provided by the highly idealized model of a
homogeneous, incompressible, non-rotating, spherical fluid body with a
viscosity proportional to the pressure (cf.\ Sridhar \& Tremaine
1992).  Let the body have mass $M$, radius $R$ and dynamical frequency
$\omega_\rmd=(GM/R^3)^{1/2}$, and let the dynamic viscosity be $\alpha
p/\omega_\rmd$, where $p$ is the pressure and $\alpha$ is a
dimensionless viscosity coefficient.  Then the linearized equations
are solved by an irrotational displacement proportional to (but out of
phase with) the tidal acceleration $-\bnabla\Psi$, and the potential
Love number can be shown to be
\begin{equation}
  k_l^m(\omega)=\f{3}{2(l-1)}\left[1-\f{\omega^2}{\omega_l^2}-\left(\f{2l+1}{l}\right)\rmi\alpha\left(\f{\omega}{\omega_\rmd}\right)\right]^{-1},
\end{equation}
which is independent of $m$ because of the spherical symmetry of the
model.  Here $\omega_l=[2l(l-1)/(2l+1)]^{1/2}\omega_\rmd$ is the
natural frequency of the f~mode of degree~$l>1$, which is the only
oscillation mode that can be excited in this simple model.  [This is
the fundamental mode or surface gravity mode, first derived by Kelvin
(Thomson 1863).]  Note that the hydrostatic response
$k_l=k_l^\mathrm{hom}=3/[2(l-1)]$ is obtained in the low-frequency
limit.  The dynamics is exactly equivalent to that of a forced damped
harmonic oscillator with natural frequency $\omega_l$ and damping
coefficient $2(l-1)\alpha\omega_\rmd$.  Now
\begin{equation}
  \imag[k_l^m(\omega)]=\f{3(2l+1)}{2l(l-1)}\alpha\left(\f{\omega}{\omega_\rmd}\right)\left[\left(1-\f{\omega^2}{\omega_l^2}\right)^2+\left(\f{2l+1}{l}\right)^2\alpha^2\left(\f{\omega}{\omega_\rmd}\right)^2\right]^{-1}.
\end{equation}
For a weakly damped oscillator ($\alpha\ll1$) this response function
has a strong peak close to the natural frequency (and attains
$|\mathrm{Im}[k_l^m(\omega)]|\gg1$, as mentioned in Section~2.2).
Well below this frequency, the factor in square brackets is close to
unity, corresponding to a frequency-dependent tidal quality factor
given by $Q^{-1}\approx[(2l+1)/l]\alpha(\omega/\omega_\rmd)$, or a
frequency-independent tidal time lag
$\tau\approx[(2l+1)/l](\alpha/\omega_\rmd)$.  (For this model, $Q'$
and $Q$ are equivalent because the body is homogeneous, while
$\hat\omega$ and $\omega$ are equivalent because the body is
non-rotating.)  Note that even a damped simple harmonic oscillator has
a frequency-dependent quality factor for forced oscillations.  In the
case of $l=2$, the low-frequency response of the body is simply a
hydrostatic spheroidal bulge with a small lag due to viscosity.  If
instead the body has a uniform kinematic viscosity $\nu$, as in the
model of Darwin (1880), then for $\omega\ll\omega_\rmd$ we have a
similar result with $\alpha$ replaced by $(2l+1)(\nu/R^2\omega_\rmd)$.

More realistic models of stars and planets support larger families of
oscillation modes, allowing more opportunities for resonances with
tidal forcing.  Oscillation modes of the Sun and other stars have been
studied in great detail for the purposes of helioseismology and
asteroseismology (e.g.\ Aerts et al.\ 2010).  Generally, this work
either neglects rotation or treats it as a perturbation: the stellar
or planetary model is spherically symmetric.  Following Cowling (1941)
the modes are classified as f (fundamental) modes (i.e.\ surface
gravity waves), p (pressure) modes (i.e.\ acoustic waves) and g
(gravity) modes (i.e.\ internal gravity waves).  The natural
frequencies of f modes and p modes ($\gtrsim10^{-4}\,\mathrm{Hz}$ for the Sun) are generally too high
($\gtrsim\omega_\rmd$) to be excited directly by tidal forcing, because
the spin and orbital frequencies are usually small compared to the
dynamical frequency of the body.  The g modes, however, which
propagate only in stably stratified (radiative) regions, form a dense
spectrum at low frequencies and are good candidates for resonant
excitation by tidal forcing.  This is the basis of Zahn's theory of
the dynamical tide (see Section~3.4 below).

As discussed in Section~2, tidal frequencies are often comparable to
spin frequencies.  This low-frequency part of the spectrum has
received little attention in asteroseismology.  The Coriolis force
cannot be regarded as a small effect, and the wave equations are not
separable in spherical harmonics.  New types of oscillation mode
(inertial waves) appear, which owe their existence to the Coriolis
force.  More interestingly, these low-frequency waves may not form
classical normal modes and the tidal response may not be describable
as a superposition of normal-mode resonances (see Section~3.5 below).

Internal waves supported by buoyancy and Coriolis forces play an
important role in the Earth's ocean and atmosphere, and have been
widely studied through theory and experiment, but are less well known
in astrophysics.  They have properties quite different from acoustic
or electromagnetic waves, being highly anisotropic and dispersive. In
the fluid frame, the wave frequency depends only on the direction, and
not on the magnitude, of the wavevector. Special directions are
defined by gravity or rotation. Energy propagates at the group
velocity, which is perpendicular to the wavevector and proportional to
the wavelength. Internal waves are highly dispersive and do not
steepen like acoustic waves, but when they exceed a critical amplitude
they break, undergoing a local instability that causes their energy to
be transferred to smaller scales and dissipated. In a closed domain,
internal waves generally do not form normal modes with regular
eigenfunctions, although the simplest cases are exceptions to this
rule.  The frequencies of internal waves are bounded and form a dense
or continuous spectrum.  The question then arises of how a rotating or
stably stratified fluid responds to a tidal force with a frequency
that lies within this spectrum.

\medskip

\textbf{3.2.\ Equilibrium and dynamical tides}

The equilibrium tide is an approximation to the tidal response of a
fluid body.  It consists of a large-scale deformation of the body, which in
the case of a quadrupolar tide ($l=2$) is a spheroidal bulge.  The
density and pressure are distorted in order to achieve hydrostatic
equilibrium in the instantaneous potential, which includes the
distorted self-gravitational potential of the body in addition to the
applied tidal potential.  This hydrostatic balance is usually a good
first approximation if the tidal frequency is small compared to the
frequency of the f~mode.  In the model considered in Section~3.1, the
equilibrium tide corresponds to the low-frequency limit $\omega\to0$
of the tidal response.

A displacement of the fluid is required in order to bring about the
tidal distortion.  However, this is not uniquely defined in general
and is a potential source of confusion or error, as discussed below.
In any case, if the tidal frequency in the fluid frame is non-zero,
the equilibrium tide does not satisfy the equation of motion exactly
because the acceleration of the fluid is neglected in computing it, as
is the effect of any damping forces.  Corrections to the equilibrium
tide generally include internal waves as well as non-wavelike
components.  Some or all of these corrections, or the complete tide
calculated without a hydrostatic approximation, have been called the
dynamical tide.

An unperturbed body in hydrostatic equilibrium satisfies
\begin{equation}
  \mathbf{0}=-\f{1}{\rho}\bnabla p-\bnabla\Phi,
\end{equation}
where $\Phi$ is the self-gravitational potential of the body, but may
also include the centrifugal potential corresponding to a uniform
rotation (in which case the body is oblate).  The curl of this
equation implies that $\bnabla\rho$ must be parallel to $\bnabla p$,
which is also parallel to $\bnabla\Phi$.  Therefore $\rho$ and $p$ are
functions of $\Phi$ only; we can also consider $\rho$ to be a function
of $p$ only.

Let the body be perturbed by a tidal potential $\Psi$ (satisfying
$\nabla^2\Psi=0$ inside the body) such that it maintains hydrostatic
equilibrium.  The linearized hydrostatic equation
\begin{equation}
  \mathbf{0}=\f{\rho'}{\rho^2}\bnabla p-\f{1}{\rho}\bnabla p'-\bnabla\Phi'-\bnabla\Psi
\end{equation}
is equivalent to
\begin{equation}
  \mathbf{0}=-\bnabla W+\left(\rho'-\f{\rmd\rho}{\rmd p}\,p'\right)\f{1}{\rho^2}\bnabla p,
\end{equation}
where $W=p'/\rho+\Phi'+\Psi$, and primes denote Eulerian
perturbations.  The curl of this equation implies that
$\bnabla(\rho'-\f{\rmd\rho}{\rmd p}\,p')$ must be parallel to $\bnabla
p$, as must $\bnabla W$.  Therefore $\rho'-\f{\rmd\rho}{\rmd p}\,p'$
must be a function of $\Phi$ only, and so must $W$.  For an
oscillatory tide (or tidal component) in which all perturbations have
zero mean, they must in fact vanish.
We deduce that
\begin{equation}
  p'=-\rho(\Phi'+\Psi),\qquad
  \rho'=-\f{\rmd\rho}{\rmd p}\rho(\Phi'+\Psi).
\end{equation}
The linearized Poisson equation then implies
\begin{equation}
  \nabla^2\Phi'=-4\pi G\f{\rmd\rho}{\rmd p}\rho(\Phi'+\Psi)
\end{equation}
inside the body, while $\nabla^2\Phi'=0$ outside.  Given $\Psi$, this
inhomogeneous Helmholtz-type equation (which can be related to the
equations of Clairaut and Radau if the unperturbed body is spherical)
can be solved, subject to suitable conditions at the surface of the
body, to determine $\Phi'$ and hence $p'$ and $\rho'$.  This is the
equilibrium tide.

More problematic, though, is the tidal displacement $\bxi$ required to
bring about these perturbations.  Assuming adiabatic perturbations, we
have
\begin{equation}
  p'=-\gamma p\bnabla\cdot\bxi-\bxi\cdot\bnabla p,\qquad
  \rho'=-\rho\bnabla\cdot\bxi-\bxi\cdot\bnabla\rho,
\end{equation}
and so
\begin{equation}
  0=\rho'-\f{\rmd\rho}{\rmd p}p'=-\left(\rho-\gamma p\f{\rmd\rho}{\rmd p}\right)\bnabla\cdot\bxi,
\end{equation}
where $\gamma=(\p\ln p/\p\ln\rho)_s$ is the adiabatic exponent.
If the body is stably stratified, having a non-zero
(Brunt--V\"ais\"al\"a) buoyancy frequency $N$, then the quantity in
brackets is non-zero and we deduce that $\bnabla\cdot\bxi=0$: the
displacement is incompressible.  It then follows that
\begin{equation}
  \bxi\cdot\bmg=\Phi'+\Psi,
\end{equation}
where $\bmg=-\bnabla\Phi=(\bnabla p)/\rho$ is the (effective)
gravitational acceleration.  This relation determines the vertical
displacement, while the horizontal displacement follows (non-locally)
from the condition $\bnabla\cdot\bxi=0$.

This solution is the standard equilibrium tidal displacement (Zahn
1966a; Goldreich \& Nicholson 1989).  It is the simplest form of
displacement that achieves hydrostatic equilibrium in the distorted
potential; under this volume-preserving motion, fluid elements retain
their density and pressure and simply move them to a different
location.  However, if the body (or part of it) is adiabatically
stratified (as in a convective zone, to a good approximation) then it
is possible to add an arbitrary `internal' displacement satisfying the
anelastic constraint $\bnabla\cdot(\rho\bxi)=0$ without changing
$\rho'$ or $p'$.  Moreover, as discussed below, such an additional
displacement is required in order to satisfy the equation of motion in
the low-frequency limit.  The total displacement then satisfies
neither $\bnabla\cdot\bxi=0$ nor $\bnabla\cdot(\rho\bxi)=0$, except in
the case of a homogeneous body.

In an adiabatically stratified region, the linearized inviscid
equation of motion for an arbitrary tidal frequency $\hat\omega$ in
the fluid frame is exactly
\begin{equation}
  -\rmi\hat\omega\bmu+2\bOmega_\rms\times\bmu=-\bnabla W,
\label{eom}
\end{equation}
where $\bmu=-\rmi\hat\omega\bxi$ is the velocity perturbation.  In the
case of a non-rotating body, or in the limit
$\hat\omega^2\gg4\Omega_\rms^2$, this equation implies that
$\bnabla\times\bmu=\mathbf{0}$ and $\bnabla\times\bxi=\mathbf{0}$: the
tidal flow and displacement are irrotational (Terquem et al.\ 1998;
Goodman \& Dickson 1998).  If we write $\bxi=\bnabla U$, then the
expression for $\rho'$ implies that
\begin{equation}
  \bnabla\cdot(\rho\bnabla U)=\f{\rmd\rho}{\rmd p}\rho(\Phi'+\Psi).
\label{irrotational}
\end{equation}
This Poisson-type equation can be solved for $U$, subject to suitable
boundary conditions, to determine $\bxi$.

However, in the limit $\hat\omega^2\ll4\Omega_\rms^2$,
equation~(\ref{eom}) implies that the tidal flow is in geostrophic
balance, and has a different form again.  If $\hat\omega^2$ is
comparable to $4\Omega_\rms^2$ then the solution generally includes
inertial waves (see Section~3.5 below).

To summarize, although the Eulerian perturbations of density and
pressure are well defined in the hydrostatic limit
$\hat\omega^2\ll\omega_\rmd^2$, the form of the tidal displacement and
velocity in the low-frequency limit depends significantly on the
ordering of $\hat\omega^2$, $N^2$ and $4\Omega_\rms^2$.  (At
sufficiently low tidal frequencies the assumption of adiabatic
perturbations may also break down.)  The standard equilibrium tidal
displacement is obtained only in the limit $\hat\omega^2\ll N^2$ and
applies only in radiative zones.

\medskip

\textbf{3.3.\ Equilibrium tides}

As described above, a time-dependent equilibrium tide is associated
with a large-scale velocity field.  Any process that resists this flow
will lead to dissipation.  Since the ordinary `molecular' viscosity is
much too small to be of interest, the main candidate for providing
linear damping of this flow is turbulent convection; nonlinear
possibilities are discussed in Section~4.1 below.  In the simplest
approach, the effect of convection on the equilibrium tide is modelled
as an effective viscosity, which can be estimated using an extension
of the same mixing-length theory that is used to describe the
convective heat flux in the theory of stellar structure.

It has long been recognized that the effective viscosity should be
reduced by some factor when the tidal period is short compared to the
typical convective timescale, as is often the case.  (In the Sun, this
timescale ranges from a month or longer near the base of the
convective zone to a few minutes near the photosphere.  In giant
planets the convective timescale is generally longer because of the
much smaller heat flux, although the applicability of mixing-length
theory is uncertain because of the rapid rotation.)  Different
reduction factors were proposed by Zahn (1966b) and by Goldreich \&
Nicholson (1977) and Goldreich \& Keeley (1977).  Zahn's reduction
factor is weaker, with $\nu\propto|\hat\omega|^{-1}$ and so
$\mathrm{Im}(k)\propto Q'^{-1}\propto|\hat\omega|^0$ for large
$|\hat\omega|$.  Goldreich's reduction factor is stronger, with
$\nu\propto|\hat\omega|^{-2}$ and so $\mathrm{Im}(k)\propto
Q'^{-1}\propto|\hat\omega|^{-1}$ for large $|\hat\omega|$.  Subsequent
theoretical arguments (Goodman \& Oh 1997; Ogilvie \& Lesur 2012)
favour this stronger reduction, which gives a frequency-dependence
similar to a Maxwellian viscoelastic model with a relaxation time
related to the convective timescale.

\begin{figure}
\centerline{\epsfysize10cm\epsfbox{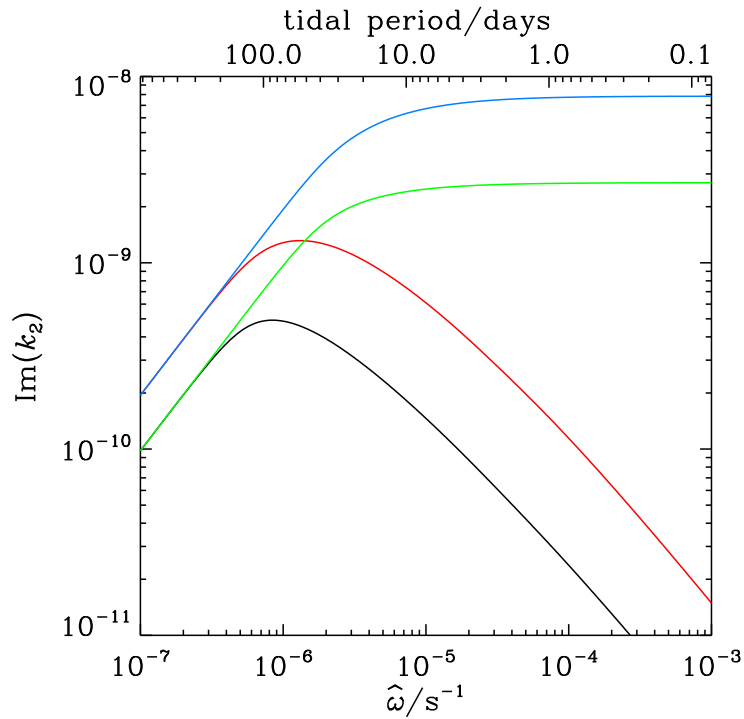}}
\caption{Frequency-dependence of the dissipation of the non-wavelike
  tide in the convective region of a standard solar model.  The
  turbulent viscosity is estimated using mixing-length theory, but is
  reduced using the prescriptions of Zahn (1966b) (blue and green
  curves) and Goldreich \& Nicholson (1977) (red and black curves) in
  the forms quoted by Zahn (1989).  The blue and red curves are
  computed assuming a standard equilibrium tide, while the green and
  black curves use the irrotational non-wavelike tide appropriate for
  a non-rotating convective region.}
\label{f:imk_sun}
\end{figure}

Fig.~\ref{f:imk_sun} illustrates the differences between these models
in the case of the solar convective zone.  The values of
$\imag[k_2^m(\omega)]$ are found by calculating the total dissipation
rate in the convective zone using the formulae quoted by Zahn (1989),
and assuming that the velocity field is either the standard
equilibrium tide or the irrotational non-wavelike tide appropriate for
a non-rotating convective region (discussed near
equation~\ref{irrotational}).  For tidal periods of a few days or
less, relevant for tidally interacting binary stars and extrasolar
planets, the reduction factors can differ by more than one order of
magnitude, and the irrotational non-wavelike tide also leads to less
dissipation.  It must be emphasized that the results for other stars,
or for the early Sun, could be very different.

In stars with deep convective envelopes, the scaling laws for the
convective timescale $t_\rmc$ and for the effective viscosity $\nu$
and resulting tidal time lag $\tau$, before any reduction factor is
applied, and omitting factors of order unity, are
\begin{equation}
  t_\rmc\sim\left(\f{M_\mathrm{env}R^2}{L}\right)^{1/3},\qquad
  \nu\sim\left(\f{LR^4}{M_\mathrm{env}}\right)^{1/3},\qquad
  \tau\sim\left(\f{LM_\mathrm{env}^2R^7}{G^3M^6}\right)^{1/3},
\end{equation}
where $M_\mathrm{env}$ is the mass of the envelope, while $M$, $R$ and
$L$ are the stellar mass, radius and luminosity.  In a star that
evolves to become a giant, $\tau$ can increase enormously, although
this phase is relatively short-lived.

More recently, numerical simulations of turbulent convection have been
used to address this issue.  The results of Penev et al.\ (2007;
2009a,b) suggest that something closer to the prescription of Zahn
(1966b) may be appropriate for intermediate frequencies, not for the
reasons originally suggested, but probably because the power spectrum
of the convection in their simulations is less steep than the
Kolmogorov spectrum assumed by Goldreich \& Nicholson (1977).
However, in a more idealized model, Ogilvie \& Lesur (2012) obtained
results that favour the steeper reduction factor and they even found
that negative effective viscosities are possible, meaning that the
convection can do work on the tidal disturbance.  Further
investigation is clearly required.

In giant planets the complex nature of the fluid may allow other forms
of dissipation.  Stevenson (1983) has argued that a large anomalous
bulk viscosity can arise in a two-phase medium such as that thought to
occur in Saturn, and possibly Jupiter, when the planet has cooled
sufficiently that helium becomes immiscible with hydrogen and forms
droplets.  The action of this bulk viscosity on the tidal flow could
provide an important amount of dissipation.  (There is some
uncertainty in Stevenson's analysis concerning the amount of
compression involved in the tidal flow.  Contrary to the argument
presented above, he states that the tidal displacement is not nearly
incompressible because the region of helium separation should be
stably stratified.)  The properties of the droplets, however, remain
very uncertain.  In the same paper, Stevenson mentions another
possibility, which is that the equilibrium tide could induce
dissipation as fluid cycles across a first-order phase transition at
the interface between molecular and metallic hydrogen.  However,
current models indicate that a first-order phase transition, although
possible in principle (McMahon et al.\ 2012), is not expected in
Jupiter (French et al.\ 2012).

A controversial suggestion by Tassoul (1987), that tidal
synchronization in close binary stars could proceed efficiently by
means of Ekman pumping involving a viscous boundary layer, was refuted
by Rieutord (1992) and Rieutord \& Zahn (1997), but resurfaces
occasionally in the literature.  In fact, Tassoul's proposed mechanism
cannot be described within the framework we have presented, or
classified as linear or nonlinear, because it produces a torque that
is of first order in the tidal amplitude parameter $\epsilon$.  This
is unphysical and apparently relies on an artificial transfer of
angular momentum across the surface of the star by a viscous torque,
rather than by gravity.  Internal viscous boundary layers could
contribute to the dissipation of the equilibrium tide, however, in
giant planets containing a solid core or other internal interfaces.

\medskip

\textbf{3.4.\ Dynamical tides in radiative regions}

Cowling (1941), who first analysed the oscillation modes of a stably
stratified (polytropic) stellar model, found that the $l=2$ g~modes
form an infinite sequence with frequencies tending to zero.  He noted
that the tidal forcing in a binary star could resonate with the higher
members of this sequence, while also suggesting that nonlinear effects
would limit the importance of such resonances.

Zahn (1970, 1975, 1977) considered the tidal forcing of low-frequency
internal gravity waves in early-type stars, which have convective
cores and radiative envelopes.  Tidal dissipation occurs because of
the radiative damping of the waves near the stellar surface, which is
more effective for lower tidal frequencies (shorter radial
wavelengths).  For sufficiently low frequencies, the waves are
strongly attenuated and do not reflect to form global g~modes.  In
this regime the tidal response scales as
$|\mathrm{Im}(k)|\propto|\hat\omega|^{8/3}$ and increases strongly
with the size of the convective core.  The dynamical tide is expected
to be more important than the equilibrium tide for higher frequencies.
The combined description of equilibrium and dynamical tides by Zahn
(1977) has been very influential (Langer 2009).

A physical interpretation (Goldreich \& Nicholson 1989) is that the
waves are launched in a narrow region near the base of the radiative
envelope, where the buoyancy frequency matches the tidal frequency and
the radial wavelength is long enough to connect to the tidal forcing.
They propagate to the surface where they are damped and deposit their
angular momentum flux.  For higher frequencies, partial reflection
occurs near the surface and the response function contains discrete
resonant peaks.  Savonije \& Papaloizou (1983, 1984) made explicit
numerical calculations, including non-adiabatic effects in full, for
particular stellar models.

A similar treatment was applied to solar-type stars, which have a
radiative core and a convective envelope, by Terquem et al.\ (1998)
and Goodman \& Dickson (1998).  Internal gravity waves are launched
near the outer boundary of the radiative core and propagate inwards.
Again, for sufficiently low frequencies, the waves are strongly
attenuated by radiative damping and
$|\mathrm{Im}(k)|\propto|\hat\omega|^{8/3}$; for higher frequencies,
the response function contains discrete resonant peaks.

Recently, related calculations have been made for carbon--oxygen white
dwarfs by Fuller \& Lai (2012b).  These stars are mostly radiative,
but may contain narrow convective zones and/or sharp features in the
profile of buoyancy frequency associated with abrupt changes in
composition.  Fuller \& Lai find that the tidal response is
approximately $|\mathrm{Im}(k)|\propto|\hat\omega|^5$, but modulated
by peaks and troughs owing to interference effects involving a wave
cavity.  It appears that the waves are mainly excited in regions where
the buoyancy frequency changes sharply, even if it remains above the
tidal frequency.

Lubow et al.\ (1997) applied Zahn's dynamical tide to hot Jupiters,
short-period extrasolar giant planets in which stellar irradiation
makes the atmosphere stably stratified to a depth of order
$100$~{bars}.  This is a promising mechanism of tidal dissipation in
such planets that requires further investigation, as the analysis has
many uncertainties related to rotation, winds, radiative damping, wave
reflection, nonlinearity, etc.  This problem would best be treated
through numerical simulations that can treat simultaneously the
nonlinear atmospheric dynamics and the tidally forced waves.
[Earlier, Ioannou \& Lindzen (1993) proposed that rotationally
modified g~modes could be excited by tidal forcing in Jupiter itself,
but their model relies on the existence of a deep, weak stable
stratification, which is not generally accepted.]

The effects of uniform rotation on the dynamical tide in radiative
regions have been examined for early-type stars by Savonije et al.\
(1995), Savonije \& Papaloizou (1997), Papaloizou \& Savonije (1997)
and Witte \& Savonije (1999a), and for solar-type stars by Savonije \&
Witte (2002) and Ogilvie \& Lin (2007).  In this work the centrifugal
distortion of the star is neglected, being of second order in the
ratio of the spin frequency to the dynamical frequency, but the
Coriolis force is retained, either in full or in the so-called
traditional approximation.  This approximation [originally introduced
by Laplace, and given its name by Eckart (1960)] retains only the part
of the Coriolis force that acts on the horizontal motion, which may be
assumed to be large compared to the vertical motion in a radiative
zone if the tidal frequency is small compared to the buoyancy
frequency.

While each oscillation mode of a non-rotating star or planet is
described by a pure spherical harmonic, and can be driven linearly
only by a tidal component of the same form, the Coriolis force in a
rotating body breaks the spherical symmetry and couples spherical
harmonics of different degrees $l$.  Retention of the full Coriolis
force, even in spherical geometry, requires two-dimensional
computations, while the traditional approximation allows a convenient
separation of variables that is analogous to, but more complicated
than, the non-rotating case.  The horizontal structure of the
separated solutions is described by Hough modes, which are
eigenfunctions of the Laplace tidal equations.  Laplace (1775; see
Cartwright 1999; Deparis et al.\ 2013) originally derived these
equations for a shallow, incompressible ocean of uniform depth on a
rotating planet, and Hough (1897, 1898) studied their solutions in
detail.  Since each Hough mode contains multiple values of $l$,
coupled by the Coriolis force, each tidal component can excite a
richer variety of modes than in the non-rotating case.

\begin{figure}
\centerline{\epsfysize5.8cm\epsfbox{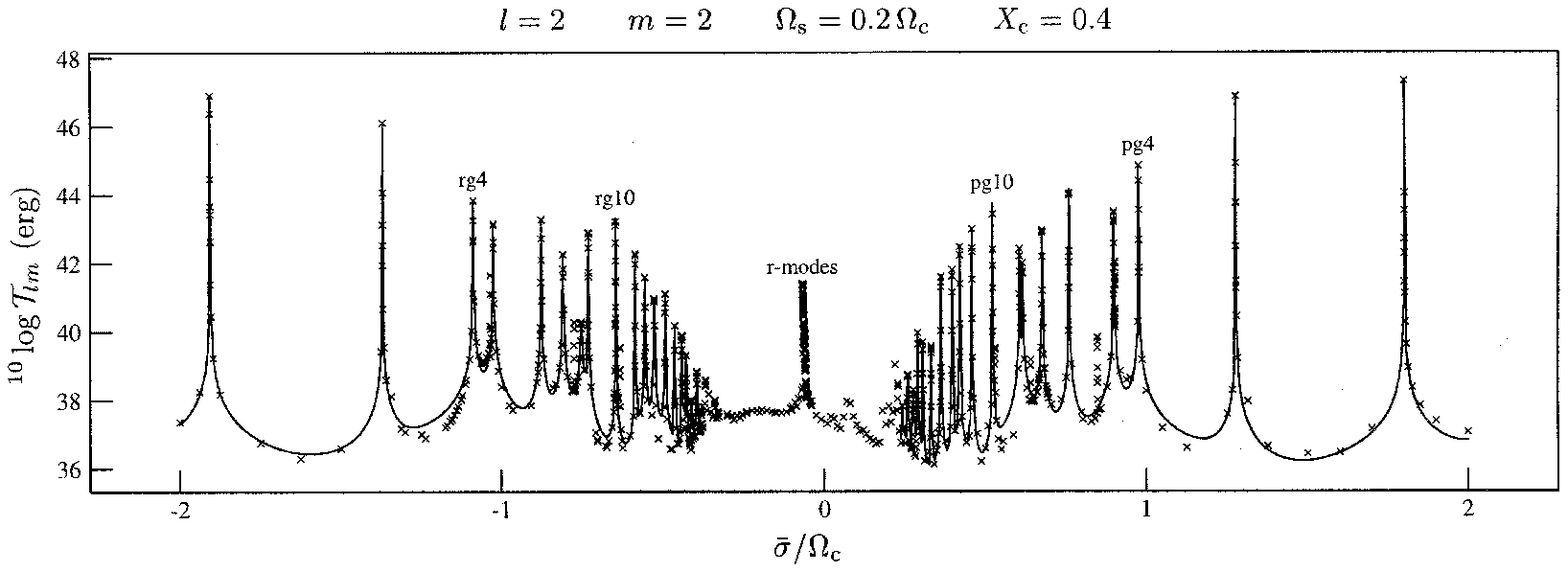}}
\caption{Linear response of a $10\,\mathrm{M}_\odot$ star, uniformly
  rotating with $\Omega_\rms=0.2\,\omega_\rmd$, to $l=m=2$ tidal
  forcing (reproduced, with permission, from Witte \& Savonije 1999a).
  In our notation, the horizontal axis is $\hat\omega/\omega_\rmd$.
  The quantity
  $|\mathcal{T}_{lm}|\approx1.3\times10^{44}|\imag(k_2^2)|$ for this
  figure.  Resonances with prograde and retrograde rotationally
  modified g~modes of different radial orders are labelled. The
  low-frequency response involves quasi-toroidal modes.}
\label{f:ws99}
\end{figure}

\begin{figure}
\centerline{\epsfysize8cm\epsfbox{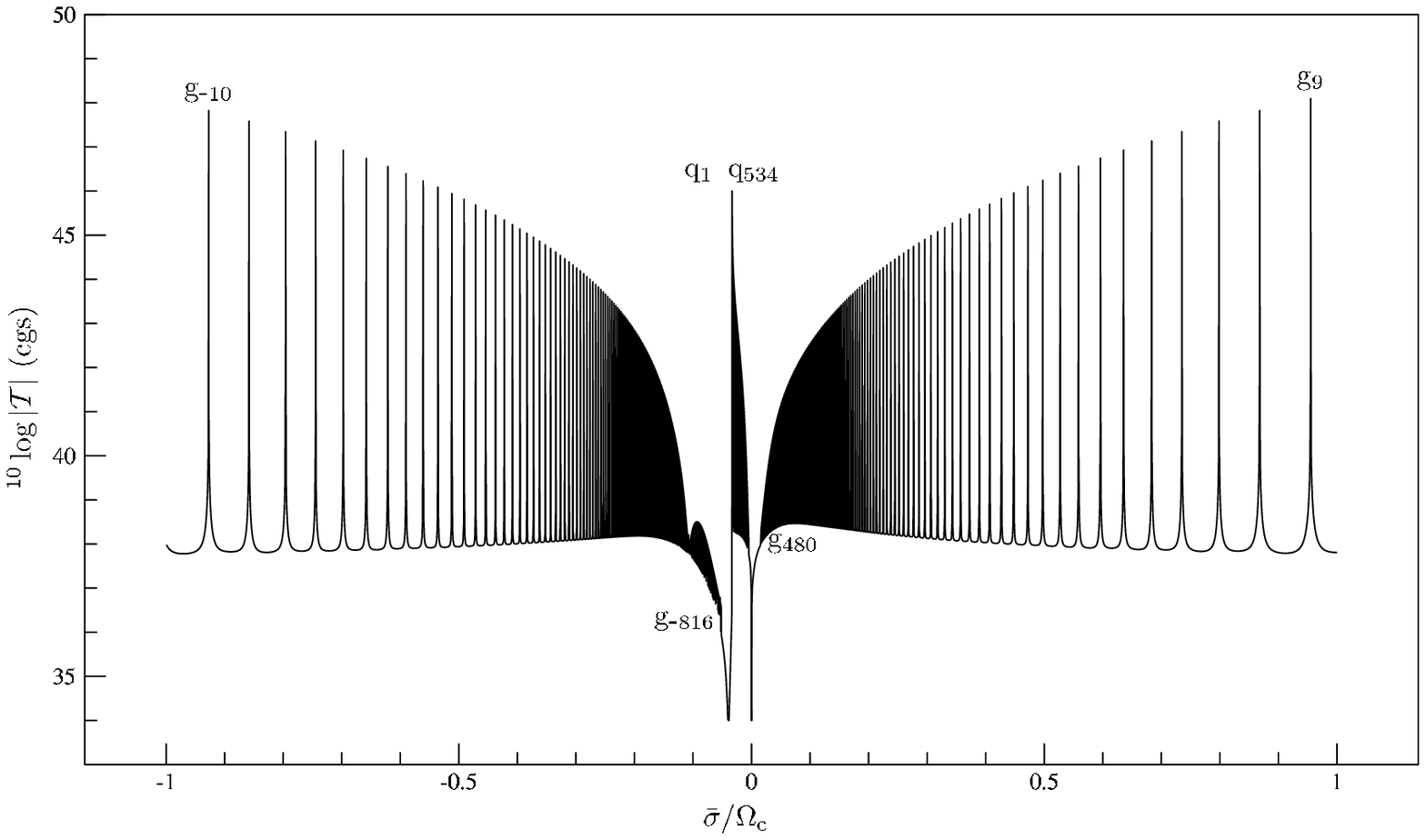}}
\caption{Linear response of a $1\,\mathrm{M}_\odot$ star, uniformly
  rotating with $\Omega_\rms=0.1\,\omega_\rmd$, to $l=m=2$ tidal
  forcing (reproduced, with permission, from Savonije \& Witte 2002).
  In our notation, the horizontal axis is $\hat\omega/\omega_\rmd$.
  The quantity
  $|\mathcal{T}|\approx6.9\times10^{44}(\mathrm{R}_\odot/R_1)|\imag(k_2^2)|$
  for this figure. Resonances with prograde and retrograde
  rotationally modified g~modes of different radial orders are
  labelled. The low-frequency response involves quasi-toroidal modes.}
\label{f:sw02}
\end{figure}

Examples of frequency-dependent linear tidal responses of uniformly
rotating stellar models are shown in Fig.~\ref{f:ws99} for
$10\,\mathrm{M}_\odot$ and in Fig.~\ref{f:sw02} for
$1\,\mathrm{M}_\odot$.  In both cases the response is dominated by
resonances with rotationally modified g~modes at higher frequencies
and with quasi-toroidal modes at lower frequencies.  These latter
modes owe their existence to the Coriolis force and are related to the
r~modes of Papaloizou \& Pringle (1978) as well as the Rossby waves
well known in geophysical fluid dynamics.

Although the calculations of Savonije \& Witte (2002) include the
convective zone, the traditional approximation breaks down there and
gives an inaccurate description of the inertial waves (Ogilvie \& Lin
2004).  Dynamical tides in convective zones are discussed in
Section~3.5 below.  The `baseline' value between resonances in
Fig.~\ref{f:sw02} can perhaps be interpreted as being due to turbulent
viscosity acting on the tide in the convective zone, although the role
of the Coriolis force in this region under the traditional
approximation is unclear. It corresponds to
$|\imag(k_2^2)|\gtrsim1\times10^{-7}$, which appears inconsistent with
Fig.~\ref{f:imk_sun}, even under optimistic assumptions.  It turns out
that Savonije \& Witte's implementation of mixing-length theory (and
of the reduction factor) means that the turbulent viscosity is much
larger than that estimated by Ogilvie \& Lin (2007), and may also be
much larger than the turbulent thermal diffusivity, contrary to
expectations.

\medskip

\textbf{3.5.\ Dynamical tides in convective regions}

Models of stellar structure treat convective zones as being very
slightly unstably stratified (i.e.\ superadiabatic), while interior
models of giant planets are usually neutrally stratified (i.e.\
adiabatic).  This is because, according to mixing-length theory, only
a very small superadiabatic gradient is usually sufficient to provide
the required heat flux through the body.

A simple approach to studying tides in convective regions is to
neglect the convection per se and to consider linear disturbances to a
neutrally stratified basic state.  If uniform rotation is also assumed
for simplicity, then the low-frequency waves that comprise the
dynamical tide are pure inertial waves restored by the Coriolis force
(e.g.\ Greenspan 1968).  If the body is fully convective then the
inertial waves propagate in a full sphere (or spheroid, more
accurately).  In other cases the waves may propagate only in a shell,
if they are excluded from either a solid planetary core or a stably
stratified fluid core that the inertial waves cannot enter directly.

Linear inertial waves in a uniformly rotating, inviscid,
incompressible fluid have been studied by Kelvin (Thomson 1880),
Poincar\'e (1885), Bryan (1889), Cartan (1922) and many others (e.g.\
Greenspan 1968).  They have remarkable mathematical properties.  Their
frequencies in the fluid frame satisfy
$-2\Omega_\rms<\hat\omega<2\Omega_\rms$, and the spatial structure of
an inertial wave with a frequency in this range is governed by a
second-order partial differential equation (named after Poincar\'e)
that is hyperbolic, with information propagating along characteristic
surfaces that are inclined at an angle
$\arcsin|\hat\omega/2\Omega_\rms|$ to the rotation axis.  Waves with a
given azimuthal wavenumber $m$ satisfy an equation with similar
properties in the meridional plane, with characteristic curves
inclined at the same angle, which is also the inclination of the group
velocity of the waves.  Hyperbolic equations with conditions on the
boundary of a closed domain generally give rise to ill posed problems
and to the formation of singularities.

The global behaviour of inertial waves in a reflective container is
determined in part by the properties of the ray circuits, which depend
sensitively on the angle of propagation, and therefore on the wave
frequency.  In a full sphere or spheroid, the rays are space-filling
for most frequencies.  It turns out that the waves can form smooth
global normal modes at certain frequencies that are dense in the
interval $(-2\Omega_\rms,2\Omega_\rms)$.  In a spherical shell,
however, for most frequencies the rays are focused towards a small
number of limit cycles known as wave attractors (Maas \& Lam 1995;
Rieutord et al.\ 2001), which exist in certain bands of frequency,
depending on the radius ratio of the shell. Generally speaking,
inertial waves do not form smooth eigenmodes in a shell, although
modes can be found localized around attractors when viscosity is
introduced.  In generalizations of this work, Dintrans et al.\ (1999)
have found curved wave attractors for gravito-inertial waves in
spherical geometry, as have Baruteau \& Rieutord (2013) for inertial
waves in a differentially rotating fluid.

In the past decade several studies have appeared examining linear
dynamical tides in such configurations, showing that the Coriolis
force gives rise to an enhanced tidal response in the frequency range
$-2\Omega_\rms<\hat\omega<2\Omega_\rms$ in which inertial waves can
propagate, and which is commonly found in applications (see
Fig.~\ref{f:frequencies}).  Ogilvie \& Lin (2004) considered a
polytrope with a solid core, and found that the tidal dissipation rate
has a strong dependence on frequency in this range.  This effect has
been explored further by Ogilvie (2009) and Rieutord \& Valdettaro
(2010).  The frequency-dependence arises because of the sensitivity of
the global propagation of inertial waves in a spherical shell to the
inclination of the rays.  It is more pronounced when the Ekman number
$\mathrm{Ek}=\nu/2\Omega_\rms R^2$ is smaller and the rays can
propagate further before dissipating.  The behaviour in the
astrophysical limit of very low Ekman number is complicated, although
there seem to be intervals of frequency in which the dissipation rate
becomes independent or weakly dependent on the Ekman number.  In a
two-dimensional version of the same problem it was shown that the
dissipation rate is asymptotically independent of the magnitude and
form of the small-scale dissipative mechanism (Ogilvie 2005); the wave
attractor simply absorbs whatever energy flux is focused towards it.
Although this result would be very appealing for astrophysical
applications, its direct applicability is limited because of the
competition between wave attractors and a singularity at the critical
latitude, where the rays are tangent to the core (Goodman \& Lackner
2009; Ogilvie 2009; Rieutord \& Valdettaro 2010).

In contrast, studies of inertial waves in a full sphere, where the ray
dynamics is much simpler, do not show this behaviour.  Instead, global
normal modes, similar to those found by Bryan (1889), are excited,
albeit rather weakly (Wu 2005a,b; Ivanov \& Papaloizou 2010).

\begin{figure}
\centerline{\epsfysize3.8cm\epsfbox{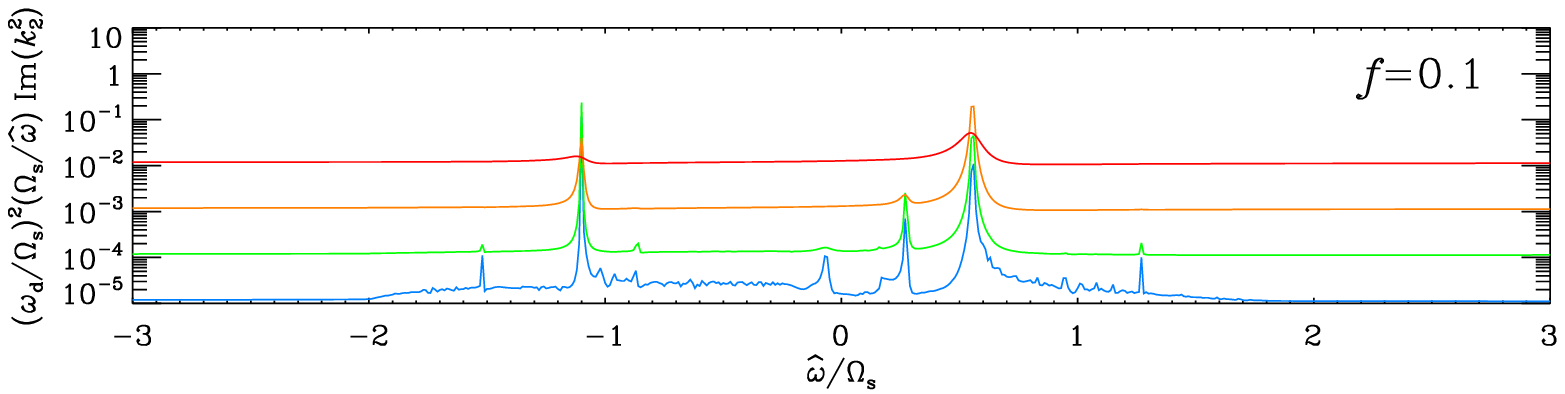}}
\centerline{\epsfysize3.8cm\epsfbox{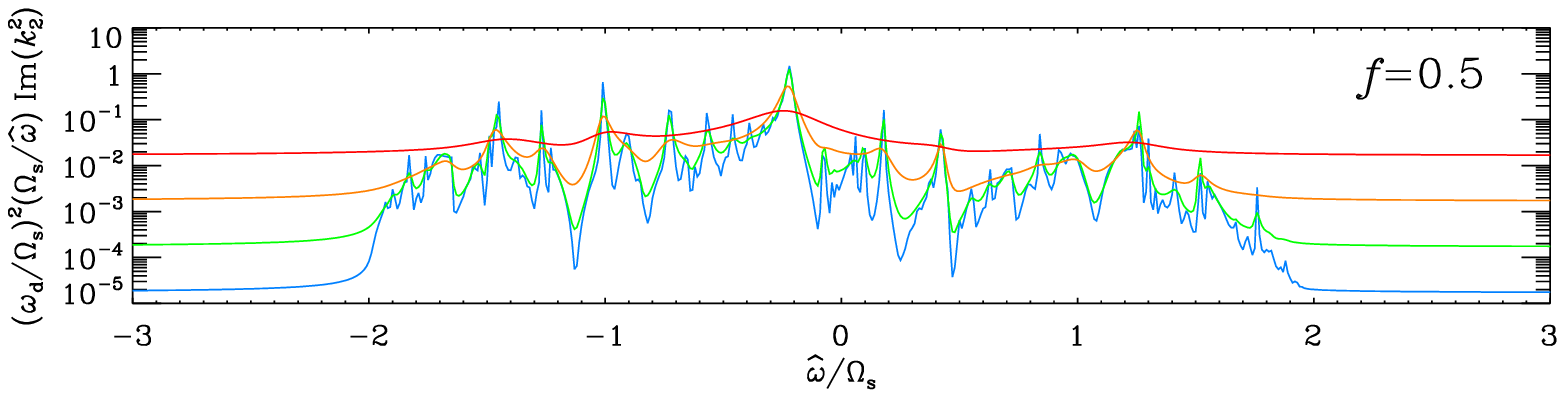}}
\caption{Linear response of a uniformly (and slowly) rotating $n=1$
  polytrope, with a perfectly rigid solid core of fractional size
  $f=0.1$ (top) or $f=0.5$ (bottom), to $l=m=2$ tidal forcing, as in
  Ogilvie (2013).  The fluid is viscous and the Ekman number is
  $\mathrm{Ek}=10^{-3}$ (red line), $10^{-4}$ (orange line), $10^{-5}$
  (green line) or $10^{-6}$ (blue line).}
\label{f:o13}
\end{figure}

\begin{figure}
\centerline{\epsfysize7cm\epsfbox{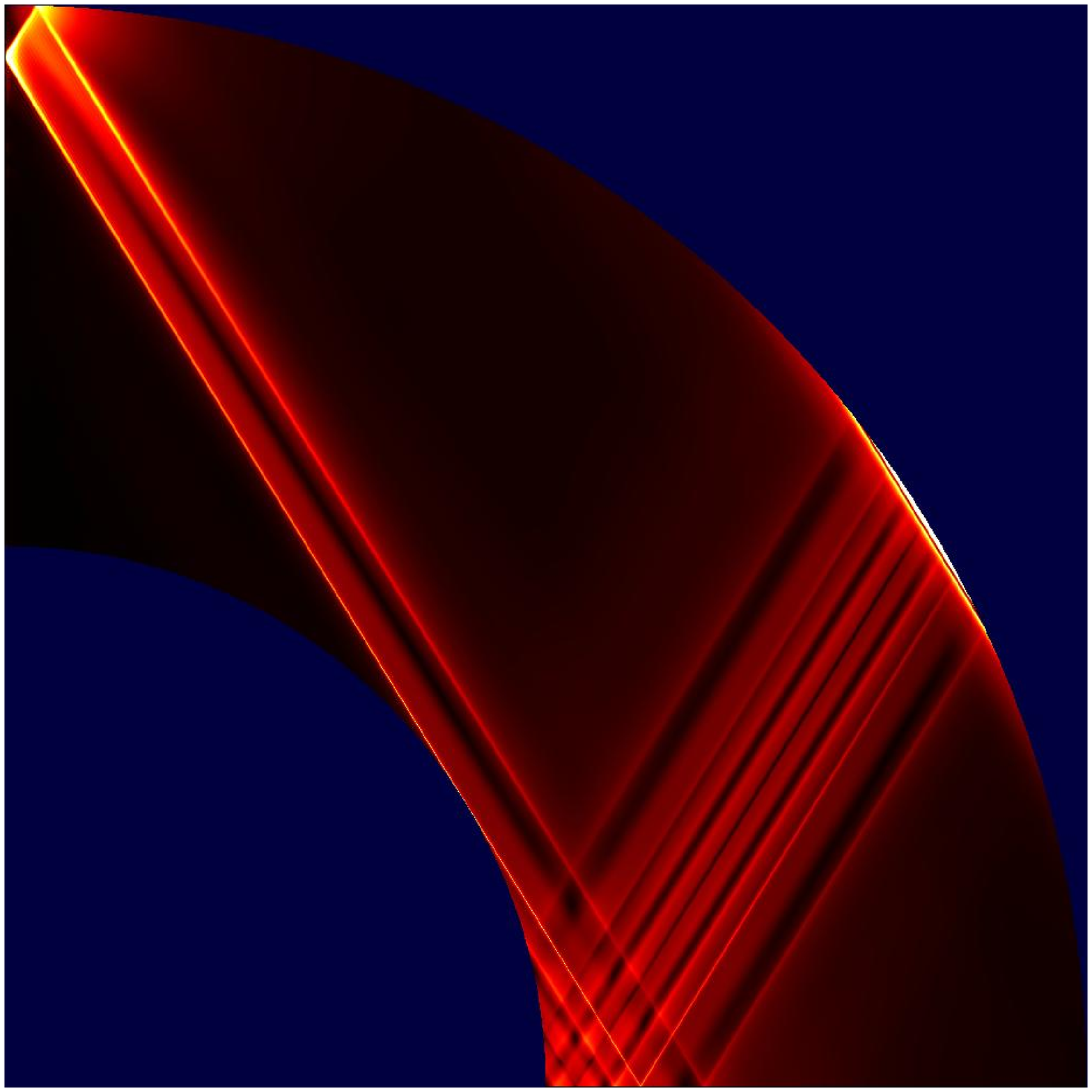}\qquad\epsfysize7cm\epsfbox{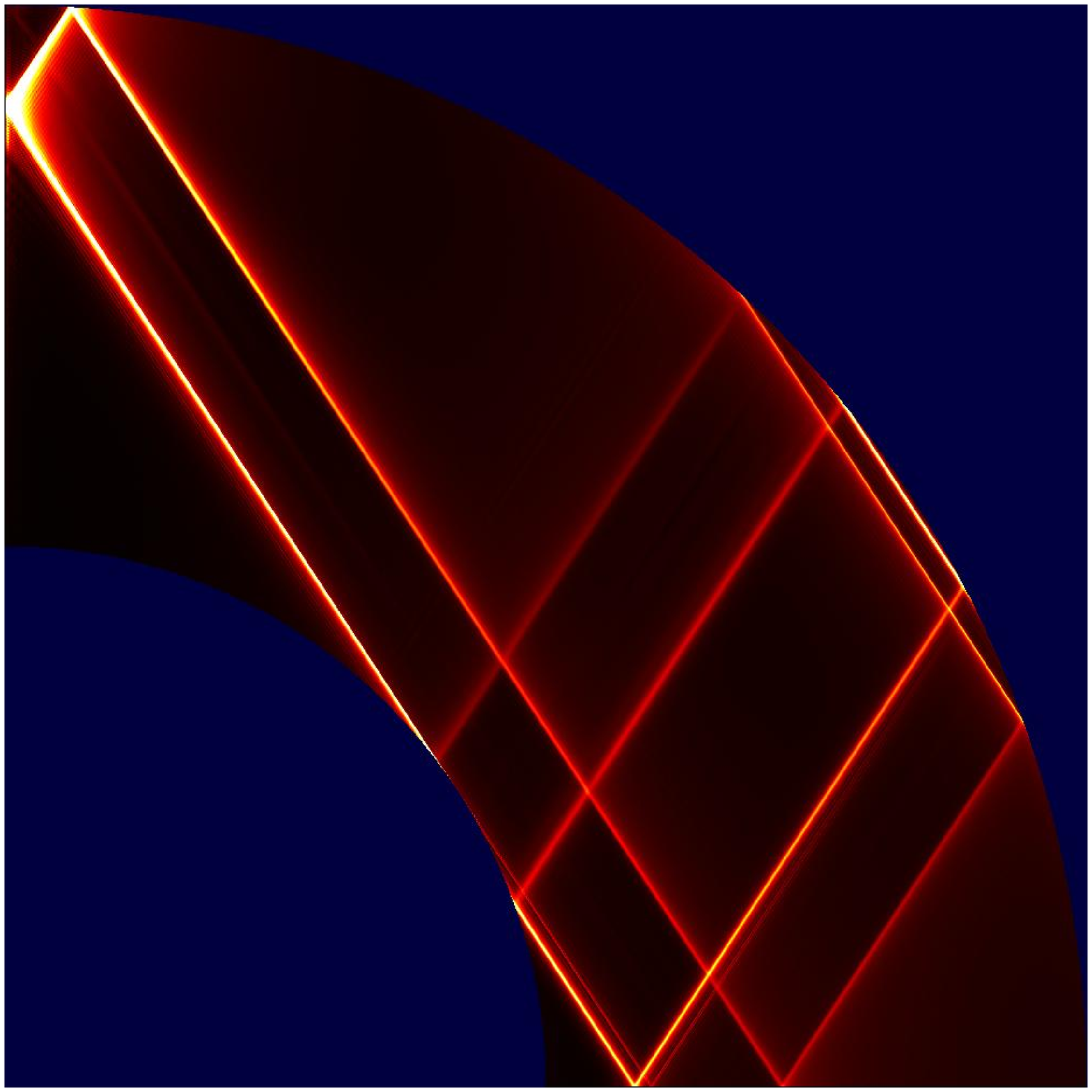}}
\caption{Structure of the velocity perturbation $|\bmu|$ in a
  meridional quarter-plane for the $l=m=2$ tide forced in a uniformly
  (and slowly) rotating homogeneous fluid body with a perfectly rigid
  solid core of fractional size $f=0.5$, as in Ogilvie (2009).  The
  fluid experiences a frictional damping force with a damping
  coefficient $\gamma=10^{-3}\,\Omega_\rms$.  In the case
  of $\hat\omega/\Omega_\rms=1.05$ (left) the response is dominated by a
  beam of inertial waves emitted from the critical latitude on the
  inner boundary.  In the case of $\hat\omega/\Omega_\rms=1.10$ (right)
  it is dominated by an inertial wave attractor.}
\label{f:o09}
\end{figure}

Examples of the frequency-dependent tidal responses are shown in
Fig.~\ref{f:o13}, for an $n=1$ polytrope with a perfectly rigid solid
core, based on Ogilvie (2013).  Responses are calculated for different
values of the Ekman number, in an effort to understand the
astrophysical limit $\mathrm{Ek}\ll1$; realistic values of
$\mathrm{Ek}$ for giant planets are beyond current computational
capabilities.  For $|\hat\omega/\Omega_\rms|>2$ the response is
uninteresting and similar to an equilibrium tide, with
$\mathrm{Im}(k)\propto\mathrm{Ek}$.  A greatly enhanced, but strongly
frequency-dependent, tidal response occurs for
$|\hat\omega/\Omega_\rms|<2$ because of the excitation of inertial
waves.  For a polytrope with a small core the response is dominated by
near-resonances with two large-scale, global inertial modes of the
full sphere, which are approximate solutions in the presence of a
small core.  When a larger core is introduced, normal modes are not
excited and the response is dominated by wave singularities associated
with the critical latitude on the inner boundary and with wave
attractors (Fig.~\ref{f:o09}; Ogilvie 2009).

A synthesis of these findings was attempted by Ogilvie (2013), who
devised a method to obtain a frequency-averaged tidal response by
means of an impulse calculation.  The quantity computed is
$\int\mathrm{Im}(k_l^m)\,\rmd\hat\omega/\hat\omega$, integrated over
the low-frequency part of the spectrum where inertial waves are found.
It can be taken as a gross measure of the tidal response in this range
of frequencies, when the details of the response curve are washed out.
In particular, for the $l=m=2$ tidal component most often considered,
the analytical result for a uniformly rotating homogeneous fluid body
with a perfectly rigid core of fractional size $f$ is
\begin{equation}
  \int\mathrm{Im}(k_2^2)\,\f{\rmd\hat\omega}{\hat\omega}=\f{100\pi}{63}\left(\f{\Omega_\rms}{\omega_\rmd}\right)^2\f{f^5}{1-f^5}.
\end{equation}
Several points are noteworthy.  First, the result is independent of
the viscosity, which is consistent with the fact that the response
curve involves more features, with higher peaks and lower troughs, as
the viscosity is reduced.  Second, the tidal response of inertial
waves increases with the square of the spin frequency, which is an
important general property (valid for $\Omega_\rms\ll\omega_\rmd$;
tidal dissipation in a rapidly rotating body has not yet been
calculated).  Third, the response vanishes for a full sphere and
increases strongly with the core size, even diverging in the limit of
a thin shell, although it should be noted that the core is assumed to
be perfectly rigid and the planet to be of uniform density.  (Tides in
shallow rotating fluid layers, first considered by Laplace, are worthy
of further investigation.)

It turns out that the dependence $f^{2l+1}$ for $f\ll1$ occurs for all
sectoral harmonics ($m=l$), while for tesseral harmonics ($m\ne l$)
the frequency-integrated response remains non-zero as $f\to0$.  This
difference in behaviour occurs because tesseral harmonics are able to
resonate with the global inertial modes of a homogeneous full sphere,
while the sectoral harmonics are not.  Care is required, though,
because the $l=2$ tesseral harmonics have only formal resonances with
the `spin-up' and `spin-over' modes, which involve rigid rotations and
have no associated dissipation, and these dominate the integrated
responses.

When the body is inhomogeneous, global modes can be excited even by
sectoral harmonics in a full sphere.  The strong dependence of the
frequency-averaged response on the core size is not seen in polytropes
of higher index that resemble centrally condensed stars.  If the solid
core is replaced by a fluid core such that the density is
discontinuous at the interface between fluid layers, then behaviour is
found that is qualitatively similar but generally weaker, especially
if the discontinuity is small.  This occurs because of the different
responses of the layers, and because the inertial waves undergo
partial internal reflection at the interface.

To summarize, as the size of the core is increased, a transition
occurs from a response involving classical resonances with global
normal modes to a richer one in which wave singularities are dominant.
Generally a greater tidal response in inertial waves is obtained by
having a larger core or a greater density contrast.  It is expected
that differential rotation will also increase the richness of the
response (Ogilvie \& Lin 2004; Baruteau \& Rieutord 2013).

\medskip

\textbf{3.6.\ Tidal encounters}

Press \& Teukolsky (1977) performed a classic calculation of the
energy deposited in the oscillation modes of a non-rotating star
during a parabolic tidal encounter.  While their study was motivated
by the binary capture mechanism of Fabian et al.\ (1975), it is also
relevant to the process by which extrasolar planets can reach
short-period orbits through tidal interaction with the star if they
start from a highly elliptical orbit.  [Errors in the numerical
results of Press \& Teukolsky (1977) were corrected by Giersz (1986),
Lee \& Ostriker (1986) and McMillan et al.\ (1987).]  An important
dimensionless parameter in the problem is, in our notation,
\begin{equation}
  \eta=\f{\omega_\rmd}{\Omega_\mathrm{max}}=\left(\f{M_1}{M}\right)^{1/2}\left(\f{d_\mathrm{min}}{R_1}\right)^{3/2},
\end{equation}
where $\Omega_\mathrm{max}=(GM/d_\mathrm{min}^3)^{1/2}$ is the
circular orbital angular velocity at the minimum orbital separation,
$d_\mathrm{min}$.  The parameter $\eta$ is a measure of the duration
of periastron passage, relative to the dynamical timescale of the
star, and is not directly related to the tidal amplitude parameter
$\epsilon$.  It is expected that $\eta\gtrsim1$ in order to avoid
tidal disruption; in the case of an incompressible fluid body with
$M_1\ll M$, tidal disruption occurs for $\eta\lesssim2.2$ (Kosovichev
\& Novikov 1992; Sridhar \& Tremaine 1992).  In a neutrally stratified
star, the energy transferred, $\Delta E$, is found to decline
exponentially with $\eta$ for large $\eta$; this is because the
orbital motion at pericentre is too slow to excite the f~mode and
p~modes of the star efficiently.  In a stably stratified star,
however, $\Delta E$ declines as an inverse power of $\eta$, because of
the excitation of low-frequency g~modes.

Lai (1997) has extended these calculations to include uniform rotation
within the traditional approximation, while Papaloizou \& Ivanov
(2005) and Ivanov \& Papaloizou (2007) have included the full Coriolis
force in the case of a coreless model with neutral stratification.
Here, global inertial modes dominate the response for large $\eta$,
and the transfers of energy and angular momentum in the tidal
encounter, $\Delta E$ and $\Delta J$, depend on the rotation rate
through the parameter $\Omega_\rms/\Omega_\mathrm{max}$, as well as on
$\eta$.  Papaloizou \& Ivanov (2010) were able to include a solid
core, even though global inertial modes do not occur, by solving an
initial-value problem for linear tidal disturbances in a parabolic
encounter.  They found that $\Delta E$ increases with the size of the
core, which is broadly in line with the findings described in
Section~3.5 for periodic tides.

Indeed, a synthesis of the results on periodic and aperiodic tides can
be achieved through the frequency-dependent response function
$k_l^m(\omega)$.  The energy and angular momentum transferred in a
parabolic encounter are equal to integrals of
$\mathrm{Im}[k_l^m(\omega)]$ over all frequencies, with two different
weightings involving the temporal Fourier transform of the tidal
potential for a parabolic orbit.  This produces a
viscosity-independent result that integrates out all the complicated
structure in the response curve.

This calculation of $\Delta E$ and $\Delta J$, which implicitly
assumes that the body is unperturbed before the encounter, is strictly
relevant for a highly eccentric orbit only if the tidal disturbance is
efficiently damped within an orbital period.  If there is interference
between waves excited by successive encounters, then the coupled
dynamics of the orbit and oscillation modes can be initially chaotic
(Mardling 1995a), although it becomes regular under the action of
dissipation (Mardling 1995b).

\medskip

\textbf{3.7.\ Discussion of linear theory}

So far we have described several situations in which linear theory
predicts a tidal response whereby $\mathrm{Im}(k)$ varies strongly as
a function of $\hat\omega$, often with peak and trough values that
differ by several orders of magnitude.  It should be borne in mind
that these response functions are based on linear theory with various
idealizations concerning the basic state.  (Even so, they are
computationally demanding because of the need for very high resolution
in the spatial and frequency domains.)  They also assume a
steady-state response; the peak values, or tidal resonances, require
multiple wave reflections and tidal periods to become established.

Nevertheless, it is important to ask how an astrophysical system
evolves in such cases.  The tidal frequency in the fluid frame for a
particular tidal component, $\hat\omega=n\Omega_\rmo-m\Omega_\rms$, is
an integer linear combination of the spin and orbital frequencies,
both of which evolve as a consequence of tidal dissipation (except in
the case of $m=0$).  The evolution of $\hat\omega$ will be most rapid at
the peaks of the response curve, and the system will therefore spend
most of its time evolving slowly through the troughs.  This would
suggest that strong tidal resonances are short-lived and of little
importance.

However, Witte \& Savonije (1999b, 2001) have found that systems can
sometimes lock into tidal resonances and undergo relatively rapid
evolution.  In general, evolution through the troughs of the response
curve of one tidal component may be accelerated or reversed by the
effects of other tidal components, or by magnetic braking or stellar
evolution, which change the shape of the response curve and the
locations of peaks and troughs.  If these other effects cause an
evolution towards a tidal resonance that itself causes an evolution in
the opposite direction, then a balance can be attained in which the
resonance is entered into to a certain degree.  Strong resonant
locking is most likely to happen when there is a near cancellation
between the rates of change of $n\Omega_\rmo$ and of $m\Omega_\rms$;
for binary stars in which the spin moment of inertia is small compared
to the orbital moment of inertia, this condition can be met for a
component with a large value of $n$, which may have appreciable
amplitude if the eccentricity is not small.

It should be borne in mind that models of stellar and planetary
interiors have been developed, for the most part, in order to
understand their gross properties and long-term evolution, not for a
detailed study of their dynamical perturbations.  In fact, stellar
interior models are increasingly being used for seismology, but this
mostly involves acoustic waves.  There are numerous uncertainties
concerning the behaviour of low-frequency waves in stellar and
planetary interiors.  In particular, it is not clear how to model
correctly the behaviour of inertial waves in a convective medium.  In
models of the Sun, for example, $-N^2$ is positive in the convective
zone and rises upwards, exceeding $4\Omega_\rms^2$ above about
$0.9\,\mathrm{R}_\odot$.  Should the linearized equations include this
unstable stratification, which would significantly alter the
properties of the waves?  Or is the unstable stratification
effectively neutralized by the existence of the convection?  Does
convection damp the waves like a viscosity, or scatter or affect them
in some other way?  What are the effects of magnetic fields,
differential rotation and imperfect reflections of the waves from the
boundaries of the convective region?

\medskip

\textbf{Sidebar: Tides in black holes}

Tides can also be raised on black holes.  Since the original work of
Hartle (1973), a number of calculations of tidal torques on black
holes have been carried out, making a variety of approximations that
ensure that the tidal interaction is weak and of low frequency
(e.g.\ Comeau \& Poisson 2009).  These are all consistent with the
viewpoint that the horizon of the black hole behaves equivalently to a
massive viscous shell with a viscous damping timescale comparable to
its light-crossing time, as in the `membrane paradigm' (Thorne et al.\
1986).  The tidal torques that have been calculated then resemble
equilibrium tides on such a shell, with an appropriate time lag
(Poisson 2009).  In reality, of course, energy and angular momentum
are exchanged through gravitational waves crossing the event horizon.
It may be expected that a richer response occurs when the tidal
frequency is comparable to that of the quasi-normal modes of the black
hole.

\medskip

\textbf{\large 4.\ Nonlinear tides}

\medskip

\textbf{4.1.\ Nonlinear equilibrium tides}

As described in Section~3.3, the equilibrium tide consists of a
spheroidal bulge that is usually time-dependent in the fluid frame and
is then associated with a large-scale velocity field.  In Section~3.3
we considered linear damping mechanisms, such as the interaction of
the tidal flow with turbulent convection, which lead to a dissipation
rate proportional to $\epsilon^2$, where $\epsilon$ is the tidal
amplitude parameter.  It is also possible that the flow could be
damped as a result of its intrinsic instability.  It would be natural
for such a mechanism to set in, or to dominate, above a critical tidal
amplitude, and to lead to a dissipation rate that is no longer
proportional to $\epsilon^2$.  This would mean that $|\mathrm{Im}(k)|$
depends on $\epsilon$, although it is then not a linear response
function but a parametrization of a nonlinear response (Goldreich
1963).

Press et al.\ (1975) noted that the Reynolds numbers of tidal flows in
close binary stars are very large.  They proposed that this leads to
turbulence and to a turbulent viscosity $\nu_\rmt$ such that the
effective Reynolds number of the tidal flow based on $\nu_\rmt$ is a
modest value of order unity, $\mathrm{Re}_\rmt$.  The resulting
scaling relations are $\nu_\rmt\sim\mathrm{Re}_\rmt^{-1}\epsilon
R^2|\hat\omega|$ and
$|\imag(k)|\sim\mathrm{Re}_\rmt^{-1}\epsilon(\hat\omega/\omega_\rmd)^2$.
However, it remains to be demonstrated whether this mechanism works in
the presence of either pre-existing turbulence (in convective zones)
or stable stratification (in radiative zones).

A more sophisticated viewpoint is that the instability would take the
form of a resonant coupling of the internal wave modes of a rotating
and/or stably stratified body.  This takes into account the fact that
the tidal flow has a well defined oscillation frequency and spatial
structure.  If two internal modes can be found with frequencies and
spatial structures that are linked by those of the tidal flow, then
that pair of modes may be destabilized.  The density of the spectrum
of internal waves makes such a coincidence more probable.  A perfectly
tuned pair of undamped modes grows at a rate proportional to
$\epsilon$, but this `parametric instability' may be reduced or
suppressed by damping or detuning of the modes.  It is much more
difficult to determine the nonlinear outcome of such instabilities and
the resulting dissipation rate, because a multitude of interacting
internal waves may be involved.

An important example is the elliptical instability, which is a
parametric resonance of internal (usually inertial) waves due to an
elliptical deformation of the streamlines of a rotating flow (Kerswell
2002).  This is precisely the situation when a spheroidal tidal bulge
is generated in a rotating body.  Elliptical instability has been
investigated by means of laboratory experiments, linear analysis and
numerical simulations, mainly for flows without stable stratification,
in which pairs of inertial waves are destabilized through their
coupling to the deformation.  Its application to astrophysical
systems, where it has sometimes been called `tidal instability', has
been discussed by Rieutord (2004), Le Bars et al.\ (2010) and C\'ebron
et al.\ (2010, 2013).  A necessary condition for elliptical
instability involving a pair of inertial waves in a uniformly rotating
body is that the primary tidal flow have $|\hat\omega|<4\Omega_\rms$
(so that it can couple to two secondary inertial waves, each having
$|\hat\omega|<2\Omega_\rms$).  For the $l=m=n=2$ tide with
$\hat\omega=2(\Omega_\rmo-\Omega_\rms)$, this condition implies
$-1<\Omega_\rmo/\Omega_\rms<3$, which is a wider range than that for
tidal forcing to excite inertial waves directly.  Even so, the
instability would not be allowed in the majority of stars hosting hot
Jupiters, where $\Omega_\rmo/\Omega_\rms>3$.

In order to observe the elliptical instability in experiments or
simulations, where the Ekman number $\mathrm{Ek}=\nu/2\Omega R^2$ is
much larger than in astrophysical bodies, the tidal amplitude must
also be made much larger than is usually realistic.  This situation
favours the dominance of large-scale modes in the nonlinear outcome of
the instability.  While attempts have been made to deduce scaling
relations, and it is claimed that an order-unity change in the basic
flow is commonly obtained, it is not yet clear what the nonlinear
outcome would be for realistic values of $\mathrm{Ek}$ and $\epsilon$.

A limitation of much of this work is that it focuses on the case in
which the deformation is stationary in the inertial frame.  This means
that the elliptical instability takes the special form of the
`spin-over' mode, which has the same property.  A typical nonlinear
outcome is a steady or recurrent tilting of the spin axis of the body,
and this has been proposed as an explanation of the spin--orbit
misalignments observed in exoplanetary systems (see Section~5.3
below).  However, a simple tilting of the stellar spin axis is not a
permissible outcome; in order to conserve the total angular momentum
of the system, the orbital angular momentum would have to increase in
magnitude, requiring an increase in the total energy of the system,
whereas in fact the total energy can only decrease as a result of
dissipation.

In an astrophysical system, where the tidal potential and deformation
rotate at the orbital frequency and the values of $\mathrm{Ek}$ and
$\epsilon$ are much smaller, it appears more likely that the
elliptical instability, if present, will generate turbulence through
the excitation of smaller-scale waves.  In a local computational
model, the instability either saturates through the formation of
long-lived columnar vortices (Barker \& Lithwick 2013), or generates
three-dimensional turbulence, in the presence of a magnetic field
(Barker \& Lithwick 2014).

Another interesting line of research has investigated nonlinear
couplings of tides and global oscillation modes (f, p and g) of
non-rotating stellar models (Kumar \& Goodman 1996; Weinberg et al.\
2012, 2013).  It provides a weakly nonlinear theory that includes
quadratic terms that couple different modes.  A conventional
application of this work is that, in radiative zones, the equilibrium
tide can excite pairs of g~modes by parametric resonance, and may
thence find a route to dissipation.  Kumar \& Goodman (1996) showed
that this mechanism could be effective in damping the f~mode excited
in a tidal encounter.  For a solar-type binary with a moderately
eccentric orbit, Weinberg et al.\ (2012) find that parametric
instability of the equilibrium tide occurs only for orbital periods
less than about one day, because the unstable modes are concentrated
near the core of the star and connect very weakly with the equilibrium
tide.  Using the same formalism, Weinberg et al.\ (2013) find a quite
different, non-resonant instability of a p--g mode pair, even for a
static (time-independent) equilibrium tide.  This result has been
challenged (Venumadhav et al.\ 2014) and appears to violate the energy
principle; it may indicate the inconsistencies of using truncated
expansions.  As yet, this approach has not included the important
effects of rotation.  The assumption of a mode expansion may be
problematic if the waves do not form global modes in the linear
regime, or if the waves are of sufficient amplitude to break.

\medskip

\textbf{4.2.\ Nonlinear dynamical tides}

In Sections~3.4 and~3.5 we discussed the tidal excitation of internal
waves and their dissipation through linear damping mechanisms.  Both
gravity waves and inertial waves may be dissipated more effectively
through nonlinear processes.  Internal waves break if they exceed a
critical amplitude, which is approximately when the displacement in
the direction parallel to gravity (or perpendicular to the rotation
axis, in the case of inertial waves) is equal to the wavelength in
that direction divided by $2\pi$.  Under these conditions, the wave so
distorts the background stratification or vorticity that a localized
convective or shear instability occurs, and turbulence may be
generated.  Below the breaking amplitude, internal waves may still be
unstable through mode couplings, and may transfer their energy to
other waves, usually of smaller scale.  In an unbounded medium or
periodic box, in the absence of dissipative effects, a plane internal
wave of any amplitude is unstable in this way (Mied 1976).

Goodman \& Dickson (1998) noted that the dynamical tide in a star
similar to the present Sun, forced by a binary companion in an
eccentric orbit, can easily exceed the breaking amplitude in the
innermost wavelengths near the stellar centre, and is then bound to be
significantly damped.  In fact, the level of nonlinearity in the wave
increases strongly towards the centre because of the concentration of
the wave energy into a small volume.  Ogilvie \& Lin (2007) applied a
similar idea to the host stars of hot Jupiters, where the stellar
rotation is slow compared to the orbit, and the eccentricity is
negligible; they found that, for a model based on the present Sun, a
companion of more than about $3$~Jupiter masses would generate waves
that could break near the centre, but also noted that the threshold
depends on the stellar model.  More detailed calculations were made by
Barker \& Ogilvie (2010), who also performed numerical simulations of
the breaking process.  Among solar-type stars, the threshold companion
mass decreases with increasing stellar mass and with increasing age.
Older or more massive stars have a steeper stratification near the
centre because of their chemical evolution.

Numerical simulations of internal gravity waves approaching the
stellar centre (Barker \& Ogilvie 2010; Barker 2011) show that, if the
breaking amplitude is exceeded, the waves are almost completely
dissipated.  Angular momentum is deposited and an expanding core is
formed in which material is spun up to the orbital frequency.  Waves
approaching the boundary of this core encounter a critical layer (at
which the angular velocity of the fluid matches that of the wave) that
causes them to be dissipated there.  The planet donates its orbital
angular momentum to the stellar core.  The resulting dissipation is
equivalent to $Q'\approx1.5\times10^5(P/\mathrm{day})^{8/3}$ for the
Sun, where $P$ is the orbital period; this estimate varies by no more
than a factor of~5 for solar-type stars between $0.5$ and
$1.1\,\mathrm{M}_\odot$.  This mechanism should lead to the
destruction of sufficiently massive hot Jupiters with orbital periods
of about one day within a few million years of the commencement of
wave breaking.  (However, if the companion is sufficiently massive,
then a tidal equilibrium may be approached instead, because the star
can be spun up to match the increasing orbital frequency; see
Section~5.3 below.)

In fact, this process is similar to one described by Goldreich \&
Nicholson (1989) in which early-type stars with close binary
companions become synchronized from the outside in, as internal
gravity waves generated by tidal forcing near the base of the
radiative envelope are absorbed by a critical layer that advances
inwards from the stellar surface.  This can be seen as a nonlinear
modification of Zahn's dynamical tide, which is more effective in
damping the waves and producing robust dissipation.

In principle, the process of an advancing critical layer could be
initiated below the breaking amplitude as waves deposit angular
momentum gradually by linear radiative damping near the stellar centre
(or surface).  However, various processes may prevent the formation of
the required differential rotation, such as hydrodynamic instability
or magnetic coupling.  It is not clear how this mechanism applies to
eccentricity tides, where there are multiple tidal components with
different angular pattern speeds (Table~\ref{t:components}).

Linear theories of dynamical tides often exhibit strong resonant peaks
in the tidal response at the frequencies of global internal modes,
where tidally forced waves reflect and interfere constructively
(Section~3.4).  If such a resonance occurs, the waves are much more
likely to break, in which case the torque will be reduced to a
non-resonant value.  The passage through such a tidal resonance could
therefore initiate an advancing critical layer.  Models that rely on
locking into a linear resonance may be compromised by the breaking of
the waves.  This was acknowledged by Witte \& Savonije (1999b), but
appears not to have been estimated accurately; Witte \& Savonije
(2002) capped their resonant torques by limiting the velocity
perturbation to the sound speed at the stellar centre, but in fact the
relevant comparison is with the much smaller wave speed of internal
(gravity) waves.

Nonlinear processes have also been discussed for inertial waves in
convective regions.  Goodman \& Lackner (2009) argued that a fraction
of the inertial wave flux generated at the critical latitude on the
inner boundary of a spherical shell would be subject to dissipation
through wave breaking.  By preventing multiple reflections within the
shell, this process would strongly reduce the frequency-dependence of
the tidal response.  Jouve \& Ogilvie (2014) have simulated an
inertial wave attractor in linear and nonlinear regimes; as waves are
focused towards the attractor, they may exceed the breaking amplitude
and produce waves of smaller scale before they themselves can reach
the scale of viscous dissipation, but the total dissipation rate is
not greatly reduced compared to the linear theory.  Favier et al.\
(2014) have performed numerical simulations of forced inertial waves
in spherical shells.  While the results for low amplitudes agree with
the linear theory, the dominant nonlinear effect they find is the
development of zonal flows, in which the angular velocity depends on
cylindrical radius in a complicated way that is related to the
locations at which the inertial waves are preferentially dissipated.
As these zonal flows build up, they increasingly alter the propagation
and dissipation of the inertial waves.  This interaction of waves and
a mean flow may be self-regulating, although in some cases
hydrodynamic instability is seen to occur.

An important nonlinear aspect of tidal interactions, therefore, is the
development of differential rotation through the non-uniform
deposition of angular momentum by tidal disturbances, and the
interaction of this differential rotation with the tides.  This adds
another layer of complexity to the problem, which is only just
beginning to be explored in detail.

\medskip

\textbf{\large 5.\ Applications to observed systems}

\medskip

\textbf{5.1.\ Direct observations of tides in stars}

Direct observations of tides in stars and giant planets are limited,
even within the solar system.  For decades, though, ellipsoidal
variations in the light curves of close binary stars have been
understood as due to the changing aspect of tidally deformed stars
(e.g.\ Kopal 1978).  Recently, Welsh et al.\ (2010) used
\textit{Kepler} to detect the much smaller (37~ppm) ellipsoidal
variations in the star HAT~P-7 due to its companion, a hot Jupiter in
a circular orbit.  \textit{Kepler} has also been used to observe
dynamical tides in several eccentric binary stars (Thompson et al.\
2012).  The light curves of these `heartbeat stars' resemble
electrocardiograms, with a brightening near periastron and persistent
oscillations throughout the orbit.  A particularly interesting example
is KOI-54 (HD~187091) (Welsh et al.\ 2011, who also mention three
pre-\textit{Kepler} cases).  This system contains two very similar
A~stars in a nearly face-on orbit of eccentricity 0.83 and period
42~days, with a minimum separation of about 6~stellar radii.  Fourier
analysis of the light curve shows that several harmonics of the
orbital frequency (notably the 90th and 91st) are preferentially
excited, as well as some non-harmonic frequencies.  Detailed dynamical
analyses of this system identify these oscillations as tidally excited
g~modes in the two stars and reveal evidence for three-mode coupling
(Fuller \& Lai 2012a; Burkart et al.\ 2012; O'Leary \& Burkart 2014).

\medskip

\textbf{5.2.\ Tidal evolution in binary stars}

Indirect evidence of tidal dissipation in stars and giant planets
comes from a comparison of the observed distributions of orbital and
spin parameters with the expected outcomes of tidal evolution, given
reasonable assumptions about the initial conditions.  For
spectroscopic binary stars, the measured joint distribution of orbital
eccentricity $e$ and orbital period $P$ provides good evidence for
tidal circularization, showing a clear preference for smaller $e$ at
small $P$.  The most detailed study is that of Meibom \& Mathieu
(2005; see also Mathieu 2005), who plot $e$ versus $P$ for eight
populations of solar-type binaries of different ages from
pre-main-sequence to late-main-sequence, and define a characteristic
circularization period (below which orbits are more circular) for each
based on a fit to the distribution.  The circularization period
generally increases with age, suggestive of ongoing tidal dissipation,
in a way that is very roughly consistent with $Q'\approx10^6$ (Ogilvie
\& Lin 2007; see also Hansen 2010, discussed in Section~5.3 below).
It is clear from Fig.~\ref{f:imk_sun} that convective damping of the
equilibrium tide in a star similar to the present Sun is much too
inefficient to explain this result, even with optimistic assumptions
($Q'\approx10^{10}$ being a more reasonable estimate).  (It should be
noted that the primary stars span a range of masses around
$1\,\mathrm{M}_\odot$, while the secondary stars, which might make an
important or even dominant contribution to circularization, are mostly
not observed.)  If the stars are assumed to be synchronized, then the
frequencies of the first-order eccentricity tides
(Table~\ref{t:components}) in the fluid frame are
$\hat\omega=\pm\Omega_\rms$ (Fig.~\ref{f:frequencies}) and can excite
inertial waves in the convective zone.  Studies of the dynamical tide
in a solar model can then explain $Q'\approx10^7$--$10^8$ (Ogilvie \&
Lin 2007), but this is still too inefficient to explain the
observations.  Witte \& Savonije (2002) found that the dynamical tide
in the radiative zone, together with the effects of rotation and
resonant locking, could also provide much more dissipation than the
equilibrium tide, but still not enough to explain the observations.  A
gap still exists, therefore, between theory and observations in this
area.  [Zahn \& Bouchet (1989) previously argued on theoretical
grounds that most of the circularization occurs during the
pre-main-sequence phase, when the stars are larger and fully
convective, but the observations do imply an ongoing process.]

Observations of stellar rotation are more difficult than those of
orbital eccentricity, and most diagnose only the rotation of the
surface.  Meibom et al.\ (2006) have measured (from periodic
photometric variability) the spin periods of the primary stars in
several solar-type binaries in two open clusters.  While they do
detect two short-period systems with circular orbits and synchronized
primaries, their other findings do not appear consistent with the
simple expectation that synchronization proceeds much more rapidly
than circularization.  A partial explanation may be that, as
synchronism is approached, the tidal frequency (in the fluid frame) of
the asynchronous tide tends to zero, while that of the eccentricity
tides tends to $\pm\Omega_\rms$, as noted above.  Therefore
$\kappa_{2,2,2}$ in equation~(\ref{sdot}) tends to zero and the
timescale for synchronization continues to increase, while the
timescale for circularization tends to a non-zero constant.

For early-type main-sequence stars with radiative envelopes,
Khaliullin \& Khaliullina (2010) have tested theories of tidal
synchronization and circularization in a set of 101 eclipsing
binaries, going beyond previous work by Claret \& Cunha (1997).
Because of concerns that the rotation of the stellar surface might be
different from that of the deep interior, they determine the rotation
periods using the observed apsidal motion together with theoretical
models of stellar structure.  After correcting an important numerical
error in Khaliullin \& Khaliullina (2007), based on a misreading of
Zahn's papers, they find that Zahn's theory of the dynamical tide is
compatible with these observations.

For giant stars, success was also found by Verbunt \& Phinney (1995),
who showed that Zahn's theory of the equilibrium tide, including the
effects of stellar evolution, could account for the circularization of
spectroscopic binaries containing a giant.  The circularization period
(more than 200~days) is sufficiently long that the controversial
reduction factor for high tidal frequencies is not relevant.

\medskip

\textbf{5.3.\ Tidal evolution in extrasolar planetary systems}

Observations of extrasolar planets by the radial-velocity method allow
measurements of the orbital period, the orbital eccentricity and the
`projected' mass of the planet, while observations of transiting
planets also provide information on the planetary radius.  In some
cases the spin period of the star can be measured and its
(sky-projected) obliquity (spin--orbit misalignment) determined from
the Rossiter--McLaughlin effect; there are no observations yet of the
spin periods or obliquities of exoplanets.  This is a rapidly
developing field that is stimulating interest in tidal interactions.
Early applications of tidal theory to exoplanets were made by Rasio et
al.\ (1996), Marcy et al.\ (1997) and Lubow et al.\ (1997).

As for binary stars, the clearest evidence for tidal interactions
between extrasolar planets and their host stars comes from the joint
distribution of orbital eccentricity and orbital period.  (Interactive
plots of up-to-date exoplanetary data can be generated at
\texttt{exoplanet.eu} or \texttt{exoplanets.org}.)  Orbits of shorter
period are more likely to be of low eccentricity (e.g.~Kipping 2013).
This can be explained by tidal circularization due to dissipation in
the planet, and is very roughly consistent with
$Q'_\mathrm{p}\approx10^{6.5}$ (Jackson et al.\ 2008; Husnoo et al.\
2012; but see also Hansen 2010, discussed below).  In some systems,
especially those with more massive planets such as CoRoT-3,
dissipation in the star may also be important for circularization.

\begin{figure}
\centerline{\epsfysize10cm\epsfbox{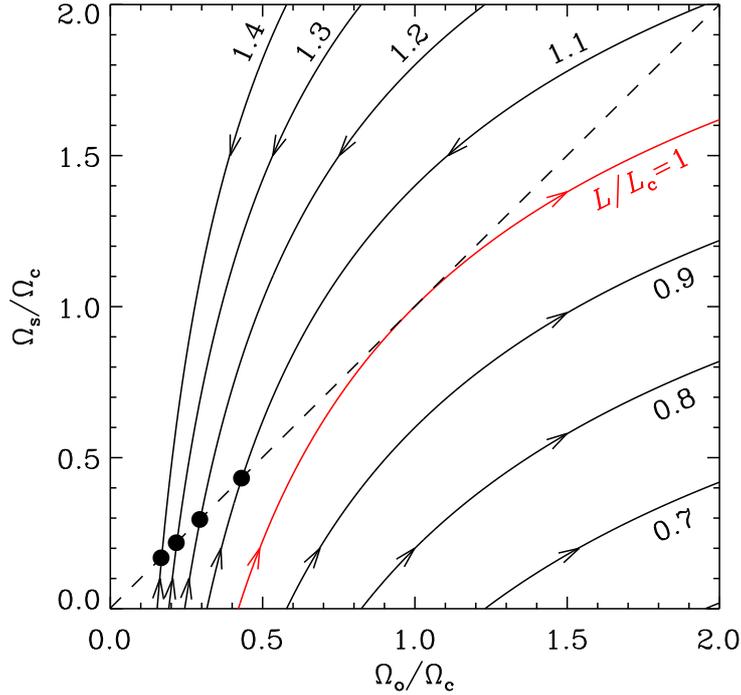}}
\caption{Evolutionary paths for a planet interacting tidally with a
  star in the circular, aligned case ($e=i=0$) in the absence of
  magnetic braking and other influences.  The axes represent the
  (stellar) spin and orbital angular velocities in units of the
  critical value (equation~\ref{lcoc}).  Contours of the total angular
  momentum $L$ are plotted, and the direction of evolution is such
  that energy is dissipated.  The planetary spin is assumed to be
  unimportant.  On the dashed line, the stellar spin is synchronized
  with the orbit.  Black points represent stable tidal equilibria when
  $L$ exceeds the critical value $L_\mathrm{c}$ (equation~\ref{lcoc}).
  Systems with $L<L_\mathrm{c}$ evolve towards planetary destruction.
  The equation of each contour is
  $(\Omega_\rms/\Omega_\rmc)+3(\Omega_\rmo/\Omega_\rmc)^{-1/3}=4(L/L_\rmc)$.}
\label{f:spinup}
\end{figure}

As noted in Section~1, in many short-period exoplanetary systems,
tidal dissipation in the star leads to inward orbital migration and
ultimately to the tidal disruption of the planet, rather than to the
synchronization of the stellar spin with the orbit as in binary stars.
Fig.~\ref{f:spinup} shows the evolutionary paths (but not the
evolutionary timescale) for a planet interacting tidally with a star
in the circular, aligned case ($e=i=0$) in the absence of magnetic
braking and other influences.  If the total angular momentum $L$ is
less than the critical value $L_\mathrm{c}$ (equation~\ref{lcoc}) then the
orbit decays and the star is spun up; if $L>L_\mathrm{c}$ then a
stable tidal equilibrium may be approached.

To interpret the observations we need theories of both planet
formation and planet destruction, and should perhaps be concerned with
the planets that are not observed as much as with those that are.
Evidence for inward migration and/or destruction is much less clear
than that for circularization.  Indeed, the existence of numerous
planets in orbits with periods of about $1$~day or less implies that
the relevant values of $Q'_*$ must greatly exceed the empirical
estimate of $10^6$ deduced from the circularization of solar-type
binary stars; otherwise the remaining lifetimes of the very
short-period planets would be implausibly short.  An extreme example
is WASP-18~b (Hellier et al.\ 2009), whose remaining lifetime would be
about $1\times10^6\,\mathrm{yr}$ if $Q'_*=10^6$; in this case the
orbital decay should be detectable within a few years through
transit-time variations (Hellier et al.\ 2009), although Watson \&
Marsh (2010) suggest that the effects could be masked by orbital
period changes resulting from a variation of the quadrupole moment of
the star during its magnetic activity cycle.  [Perhaps surprisingly,
planets of shorter orbital period such as KOI-1843.03 (Rappaport et
al.\ 2013) and Kepler-78~b (Sanchis-Ojeda et al.\ 2013) provide a
weaker constraint on $Q'_*$ because of the lower planetary masses and
higher stellar densities in those systems.]  By re-examining the
system OGLE-TR-56~b, Adams et al.\ (2011) have constrained its period
derivative to be $\dot P=-2.9\pm17\,\mathrm{ms}/\mathrm{yr}$, i.e.\
consistent with zero.  Their analysis of this result is confusing; in
fact, at the $3\sigma$ confidence level ($\dot
P>-54\,\mathrm{ms}/\mathrm{yr}$), it implies that $Q'_*>1\times10^5$.

Despite various attempts to estimate or constrain $Q_*'$ by
statistical modelling of the observed orbital semi-major axes and
eccentricities, either of populations or of single systems (Jackson et
al.\ 2008, 2009; Hansen 2010; Schlaufman et al.\ 2010; Brown et al.\
2011), there is as yet no convincing evidence of any orbital migration
due to tidal dissipation, because the initial conditions are poorly
constrained.  However, the host stars of several hot Jupiters are said
to be spinning faster than expected (Pont 2009; Husnoo et al.\ 2012),
suggesting a tidal transfer of angular momentum from the orbit to the
stellar rotation in these systems, and two stars ($\tau$~Boo and
CoRoT-3) show rotation compatible with synchronization with the orbits
of their hot Jupiters (as do KELT-1 and CoRoT-15 with their brown
dwarf companions).  As mentioned in Section~1, these apparently
synchronized systems probably have sufficient total angular momentum
to attain a tidal equilibrium.  It is not clear what dissipation
mechanism can bring this about within the age of the system, though,
because these are F~stars with shallow convective zones.

Let us recall that $Q'$ is not a fundamental property of a body; even
in linear tidal theory, it is expected to depend on the tidal
frequency as well as on the integers $l$ and $m$ that label the
different spherical harmonics.  Even though theory does not yet
explain why $Q'_*\approx10^6$ for the circularization of solar-type
binary stars, it can plausibly explain why $Q'_*$ should be much
larger for the orbital decay of hot Jupiters.  The host stars involved
are mostly spinning more slowly than those in close binary stars, and
the tidal frequency of the asynchronous tide in the fluid frame is
usually such that $|\hat\omega|\gg\Omega_\rms$.  But the eccentricity
tides in a synchronized binary star have $|\hat\omega|=\Omega_\rms$
and can excite inertial waves in the convective zone, producing a much
stronger tidal response (Ogilvie \& Lin 2007).  In the radiative zone
of a solar-type star, the asynchronous tide can break and produce
rapid orbital decay only if the planet and star are sufficiently
massive, and the star sufficiently old (Barker \& Ogilvie 2010; see
Section~4.2).  The predictions of the wave-breaking theory appear to
be in fair agreement with observations (Guillot, Lin \& Morel, in
preparation).

The most detailed comparison to date between (highly simplified) tidal
theories and exoplanetary observations is that of Hansen (2010, 2012).
He evolves interior models of the star and planet, including the
effects of tidal heating, together with the equations of tidal
evolution of the spin and orbit under the assumption of a common time
lag $\tau$, which allows arbitrary eccentricities to be considered.
In the first paper (Hansen 2010) he makes the (arbitrary) assumption
that all the stars, and all the planets, have fixed values of the
parameter (Eggleton et al.\ 1998)
\begin{equation}
  \sigma=\f{2}{3}k_2\tau\f{G}{R^5}=\f{1}{Q'|\hat\omega|}\f{G}{R^5}.
\end{equation}
The value of $\sigma_*$ that he fits from the circularization of
solar-type binary stars (see Section~5.2), which corresponds to
$Q'_*=10^6$ in the case of the present Sun for a tidal period of
$6$~days, would cause much too rapid an evolution of exoplanetary
systems, and he finds it necessary to scale it down by more than two
orders of magnitude.  His preferred value of $\sigma_\rmp$
corresponds to $Q'_\rmp=10^8$ in the case of Jupiter for a tidal
period of $4$~days, or to $10^7\lesssim Q'_\rmp\lesssim10^8$ for the
circularization of hot Jupiters.

In the second paper (Hansen 2012) he instead calculates the value of
$\sigma$ for each star or planet using Zahn's theory of the
equilibrium tide, admitting also an empirical correction factor
$\nu_0$ for the turbulent viscosity.  He finds that a mild correction
$\nu_0\approx1.5$ is sufficient to explain the circularization of
binaries containing a giant star (Verbunt \& Phinney 1995) while also
being compatible with the non-zero orbital eccentricities of several
massive, short-period exoplanets.  In fact, as illustrated in
Fig.~\ref{f:imk_sun}, Zahn's theory is not expected to be applicable
to the high tidal frequencies encountered in the latter systems.

Schlaufman \& Winn (2013) find evidence for planet destruction by
inward migration due to tidal dissipation in subgiant stars.  The
paucity of planets with $P\lesssim200$~days around subgiants, in
comparison with their presumed main-sequence progenitor stars, is
taken as evidence that the shorter-period planets were consumed as the
star swelled and developed a deep convective envelope after leaving
the main sequence.  However, it is difficult to explain the absence of
planets out to $200$~days by this means unless $Q'_*$ is very small,
of order $10^2$ (see also Hansen 2010, 2012).

Measurements of (sky-projected) stellar obliquities provide
interesting constraints on scenarios for planet formation, and tidal
evolution may play a role in their interpretation.  Surprisingly many
systems with hot Jupiters have large spin--orbit misalignments,
exceeding $90^\circ$ in some cases.  Winn et al.\ (2010) argued that
most of the cooler stars ($T_\mathrm{eff}<6250\,\mathrm{K}$) are
aligned, while the hotter stars have widely spread obliquities.  They
suggested that hot Jupiters are formed in such a way that the stellar
obliquities are broadly distributed, but those of the cooler stars are
subsequently reduced by tidal dissipation.  Albrecht et al.\ (2012)
compiled a set of 53~observed systems and argued that the aligned
systems are those in which the timescale for spin--orbit alignment is
shorter than the age.  They used the estimates of Zahn (1977) for
tidal synchronization timescales due to equilibrium and dynamical
tides, for cooler and hotter stars, respectively.  (The alignment
timescale might be assumed to be similar to the synchronization
timescale in systems for which $L_\rms\lesssim L_\rmo$; see
equations~\ref{sdot} and~\ref{idot}.)  In the case of a planet of
$1$~Jupiter mass in a $3$-day orbit around a star of
$1\,\mathrm{M}_\odot$ that is slowly rotating with a period of
$30$~days, the alignment timescale can be estimated roughly as
$2000\,Q'_*$ years, although several different tidal components are
potentially involved.  Albrecht et al.'s figure~24 (in which they
confusingly reduced the timescales by a factor of $5\times10^9$)
suggests, in fact, that tidal alignment is possible within the age of
the system for only one or two of the stars, using Zahn's theory of
the equilibrium tide.

If stellar tidal dissipation is effective in reducing the obliquity,
then there is a danger that the planet will also undergo orbital decay
and be destroyed.  From equations~(\ref{adot}) and~(\ref{idot}) and
their generalizations to arbitrary obliquity (e.g.\ Lai 2012) we can
see two situations in which alignment can proceed more rapidly than
orbital migration.  First, if $L_\rms\ll L_\rmo$, then alignment can
be faster because it involves mainly a change in the stellar spin
axis, while the orbit is hardly affected.  Second, if the tidal
response to one or more of the $n=0$ components (which are associated
with a torque but no power, and so do not affect $a$) is much
stronger, then alignment can be accelerated.  Indeed, Lai (2012) has
noted that the $2,1,0$ obliquity tide has frequency
$\hat\omega=-\Omega$ in the fluid frame, within the spectral range of
inertial waves, and may thereby elicit a much stronger tidal response.
However, Lai (2012) and Rogers \& Lin (2013) point out that boosting
this component could produce retrograde or even polar orbits, as well
as aligned prograde orbits, because the associated $\rmd i/\rmd t$ is
proportional to $-\sin i\cos^2i[1+(L_\rmo/L_\rms)\cos i]$.  Another
possible objection is that there is an $m=1$ mode at exactly
$\hat\omega=-\Omega$, but it is the spin-over mode rather than a
genuine inertial mode, and has zero dissipation because it involves a
uniform rotation.  However, there may be a non-trivial dynamics
involving a core, or involving precessional motion of the spin and
orbit due to this secular component of the tidal potential.  Further
investigation is required.

Transit observations reveal a wide spread of planetary radii, many of
which are too large to be explained by conventional models.  In
principle tidal dissipation in the planet, due to the synchronization
of its spin and the circularization of its orbit, can deposit a very
large amount of heat, and the planet can be significantly inflated if
the deposition is sufficiently deep.  However, this effect does not
persist long after the orbit is circularized.  Leconte et al.\ (2010) and
Hansen (2010) have concluded that tidal heating cannot provide a
general explanation of the radius anomalies.

Nevertheless, Arras \& Socrates (2010), following earlier studies of
Venus by Gold \& Soter (1969) and Correia et al.\ (2003), have argued
that a thermal tide, caused by asymmetric irradiation of the planet by
the star, causes a torque that drives the planet away from synchronous
spin.  The competition between this torque and the gravitational tidal
torque could result in a sustained asynchronous state with ongoing
tidal dissipation.  Although the theory of thermal tides in giant
planets is in an embryonic (and controversial) state, Socrates (2013)
finds that the expected scaling laws for the resulting tidal heating
appear compatible with the observed radius anomalies.

As well as understanding the statistical distributions of properties
of extrasolar planets, it is important to consider systems on an
individual basis, because of the varying interior structures of stars
and planets.  Future work ought to take into account the dependence of
the tidal response on the extent of a stellar convective zone, for
example, or on the size of a planetary core, where this is constrained
by observations of the planetary mass and radius.

\medskip

\textbf{5.4.\ Tidal evolution in the solar system}

\medskip

Tidal dissipation in the giant planets of the solar system leads to
orbital migration of their regular satellites, which have nearly
circular and equatorial orbits.  Except for the small, innermost moons
that orbit within the planet's corotation radius (synchronous orbit),
the migration is outward and its rate is proportional to the mass of
the satellite and to the imaginary part of the relevant Love number,
$\kappa_{2,2,2}$, at the appropriate tidal frequency.  Even without
considering the frequency-dependence of the Love number, there is a
strong dependence of the migration rate on the orbital separation or
period (see equation~\ref{adot}).  Goldreich (1965) proposed that
convergent differential migration could cause pairs of satellites to
enter the stable commensurabilities (mean-motion resonances) that are
observed in numerous cases, including the famous Laplace resonance
between Io, Europa and Ganymede.  Two or more satellites that are
locked into resonance migrate at a rate determined by their combined
angular momentum and the combined tidal torque that the planet exerts
on them.  Details of the various resonances reveal many interesting
puzzles and constraints on tidal dissipation in both the planets and
their satellites (Peale 1999).

Goldreich \& Soter (1966) deduced lower bounds on the planets'
modified tidal quality factors, $Q'=3/(-2\kappa_{2,2,2})$, from the
existence of close satellites (outside corotation), assuming they are
as old as the solar system.  It is assumed that $Q'$ does not depend
significantly on frequency.  The solution for $a$, starting from
$a=a_0$ at $t=t_0$, is then
\begin{equation}
  a^{13/2}=a_0^{13/2}+\f{117}{4}\f{M_2}{M_1}\f{R_1^5}{Q'}(GM)^{1/2}(t-t_0).
\end{equation}
In the case of Jupiter, to move Io alone from the corotation radius to
its present location in $4.5\times10^9\,\mathrm{yr}$ requires
$Q'_\mathrm{J}=1.2\times10^6$.  Once Io, Europa and Ganymede are
locked into resonance, however, their combined angular momentum is
$4.3$ times that of Io alone, and the combined tidal torque on them is
scarcely larger than that on Io alone.  This means that
$Q'_\mathrm{J}$ could be as small as $2.7\times10^5$ if the resonance
is formed early.  Combining a similar constraint with a detailed study
of the Laplace resonance and the heating of Io, Yoder \& Peale (1981)
concluded that $2.4\times10^5\lesssim
Q'_\mathrm{J}\lesssim8\times10^6$.

In the case of Saturn, to move Mimas alone from the corotation radius
to its present location in $4.5\times10^9\,\mathrm{yr}$ requires
$Q'_\mathrm{S}=8.0\times10^4$.  Once Mimas and Tethys are locked into
resonance, however, their combined angular momentum is $22$ times that
of Mimas alone, and the combined tidal torque on them is $18$ times
that on Mimas alone (assuming $Q'_\mathrm{S}$ independent of
frequency).  This means that $Q'_\mathrm{S}$ could be as small as
$6.6\times10^4$ if the resonance is formed early.

In the cases of Uranus and Neptune, the existence of Ariel and Proteus
provides lower bounds of $Q'_\mathrm{U}\gtrsim9.1\times10^4$ and
$Q'_\mathrm{N}\gtrsim6.7\times10^4$.  Based on detailed evolutionary
scenarios for Miranda, Ariel and Umbriel, Tittemore \& Wisdom (1990)
concluded that $1.6\times10^5\lesssim
Q'_\mathrm{U}\lesssim5.6\times10^5$, and most probably towards the
lower end of this range.

Recently, however, these constraints have been challenged.  Lainey et
al.\ (2009) have fitted a sophisticated dynamical model, including
parametrized tidal dissipation, to astrometric observations of the
Galilean satellites since 1891, and they deduce
$(k_2/Q)_\mathrm{J}=(1.1\pm0.2)\times10^{-5}$, i.e.\
$Q'_\mathrm{J}=(1.4\pm0.3)\times10^5$, for the asynchronous tide due
to Io.  This value, which represents only a `snapshot' of the tidal
evolution, is below the lower bounds quoted above for the average over
astronomical time and might suggest that $Q'_\mathrm{J}$ is in fact
frequency-dependent.

Applying a similar analysis to Saturn, Lainey et al.\ (2012) deduce
$(k_2/Q)_\mathrm{S}=(2.3\pm0.7)\times10^{-4}$, i.e.\
$Q'_\mathrm{S}=(7.2\pm2.2)\times10^3$, for the asynchronous tides due
to Enceladus, Tethys, Dione and Rhea.  This value is far below the
lower bound quoted above, based on the migration of Mimas (and
Tethys), and could help to explain the heating rate of Enceladus (cf.\
Meyer \& Wisdom 2007).  However, this result is controversial and
Mimas is especially problematic in the model of Lainey et al.\ because
it is apparently migrating inwards and would require a large negative
torque corresponding to $Q'_\mathrm{S}=-570$.  This suggests that the
dynamical solution of Lainey et al.\ may not be correct.  The most
plausible origin of a negative torque on Mimas's orbit is Tethys, with
which it is resonantly locked.  It is possible that a non-uniform
tidal evolution, due to a strong frequency-dependence of
$Q'_\mathrm{S}$, could cause the resonant torque between Mimas and
Tethys to fluctuate in such a way that Mimas is currently being
repelled from Tethys.

The unexpectedly rapid orbital migration deduced by Lainey et al.\
(2012) has been taken as support for a model in which planetary
satellites are formed at the outer edge of a planetary ring system
(close to the Roche limit) and migrate outwards through a combination
of ring torques and planetary tidal torques.  Initially, Charnoz et
al.\ (2010) proposed this for the small inner moons (Atlas--Janus) of
Saturn.  Charnoz et al.\ (2011) extended this idea to Mimas--Rhea,
requiring a small value of $Q'_\mathrm{S}$ as suggested by Lainey et
al.\ (2012).  (Beyond the 2:1 Lindblad resonance with the outer edge
of the rings, only the planetary tidal torque is effective.)  However,
it has not been demonstrated that the current configuration can be
reached while avoiding other stable commensurabilities such as 3:2.
More recently, Crida \& Charnoz (2012) have applied the same idea to
the other planets except Jupiter.  For Uranus and Neptune, however,
the Roche limit (however reasonably defined) is well inside the
corotation radius and the mechanism may not work; indeed, the inner
moons migrate inwards and may be disrupted to form rings (Leinhardt et
al.\ 2012).

It is natural to ask what range of values of $Q'_\mathrm{S}$ can be
explained theoretically, and whether $Q'_\mathrm{S}$ should be smaller
than $Q'_\mathrm{J}$ by an order of magnitude or more, as required to
cause a significant migration of Saturn's inner moons.  The frequency
of the asynchronous tide satisfies $|\hat\omega/\Omega_\rms|<2$ for
$0<\Omega_\rmo/\Omega_\rms<2$ (Fig.~\ref{f:frequencies}), which is
true for almost all regular satellites.  Inertial waves can therefore
be excited in the planet.  The tidal response can be expected to
depend strongly on the tidal frequency and on the size of the core, or
on other features of the internal structure.  Despite considerable
uncertainty in the interior models, values of $Q'$ of the right order
of magnitude to provide significant tidal evolution of Io and Mimas
can certainly be obtained (Ogilvie, in preparation), and Jupiter's
smaller core could explain the difference between the planets.
(Models of Uranus and Neptune are even more uncertain at the present
time.)  An alternative model involving viscoelastic dissipation in the
planetary core (Dermott 1979; Remus et al.\ 2012) could potentially
also produce values of $Q'$ of the right order (and strongly
increasing with core size, but only weakly dependent on frequency),
although the rheological parameters are very uncertain.

We should not be surprised that the values of $Q'$ for giant planets
inferred from the orbital migration of their satellites are smaller
(perhaps much smaller) than those inferred from the orbital
circularization of hot Jupiters, because the tidal frequencies are
very different in the two situations.  Furthermore, hot Jupiters are
expected to be more slowly rotating because of tidal synchronization,
and their interiors may differ in an important way from Jupiter and
Saturn; the tides in extrasolar planets are also of much larger
amplitude and may be in a more nonlinear regime.  In both cases the
tidal frequency is expected to lie in the spectral range of inertial
waves.  We have seen that the frequency-averaged response in inertial
waves scales such that $Q'^{-1}\propto(\Omega_\rms/\omega_\rmd)^2$,
which naturally provides a difference of about two orders of magnitude
between the two situations.  The alternative Maxwellian viscoelastic
model could potentially explain a difference of about one order of
magnitude between the two situations (Storch \& Lai 2014), but it also
requires that the Maxwell frequency of the core happen to be
comparable to the tidal forcing frequencies due to satellites.

\bigskip

\centerline{SUMMARY POINTS}

\begin{enumerate}
\item The irreversible evolution of the spin and orbital parameters of
  an astrophysical fluid body with a close and massive companion is
  related (at least in the case of uniform rotation) to the rate of
  dissipation of energy associated with the response of the body to
  various components of the tidal potential.
\item The response of a fluid body to tidal forcing generally consists
  of a quasi-hydrostatic bulge and an associated large-scale flow,
  together with internal (gravity and inertial) waves, often of much
  smaller scale.  Tidal forcing frequencies very often coincide with
  those of inertial waves.
\item The non-wavelike part of the tide can be dissipated by its
  interaction with convection or, in some cases, as a result of its
  own instability, or possibly through the action of irreversible
  effects in multi-phase fluids.
\item The wavelike part of the tide has a very different character in
  radiative and convective zones.  Dissipation is effective when the
  waves achieve a sufficiently small scale to be damped by radiative
  diffusion or (turbulent) viscosity, when the wave amplitude is
  enhanced by resonances with global modes, or when nonlinearity
  causes the waves to break.
\item Idealized calculations of linear tidal responses show a greatly
  enhanced, but highly frequency-dependent, dissipation rate owing to
  the excitation of internal waves.  Non-ideal linear effects and
  nonlinearity may tend to smooth out the features in the response
  curves.  The evolution of a system with a highly frequency-dependent
  tidal response is complicated and may involve resonance locking.
\item Tidal dissipation induces differential rotation in fluid bodies,
  which in turn affects the properties of the internal waves in a way
  that remains to be fully understood.
\item Observations provide indirect evidence of tidal evolution in
  binary stars, extrasolar planetary systems and the satellite systems
  of the giant planets in the solar system.  In each case useful
  constraints on theories of tidal dissipation are obtained.
\end{enumerate}

\bigskip

\centerline{FUTURE ISSUES}

\begin{enumerate}
\item Further studies of the interaction between oscillatory tidal
  disturbances and turbulent convection are needed to clarify the role
  of stellar and planetary convection in tidal dissipation.
\item Linear calculations of the tidal responses of stars and planets
  should be carried out for a wide range of realistic interior models,
  including the effects of rotation.
\item Nonlinear aspects of both wavelike and non-wavelike tides should
  be studied in more detail and under more realistic conditions.
\item The role of differential rotation, and its limitation by various
  processes inside stars and giant planets, needs to be assessed.
\item Atmospheric tides, both gravitational and thermal, in exoplanets
  should be studied using specialized numerical simulations.
\item Future observations, especially of transiting exoplanets, are
  likely to provide improving constraints on theories of tidal
  dissipation and will require systems to be modelled on an individual
  basis, taking into account their probable internal structure and
  evolutionary history.
\end{enumerate}

\bigskip

\textbf{Acknowledgements}: I am grateful to Adrian Barker, Harry
Braviner, Dong Lai, Doug Lin, John Papaloizou and Stan Peale for
comments and discussions that led to improvements in this article.

\bigskip

\centerline{GLOSSARY}

\begin{itemize}
\item \textit{body~1}: the body in which the tide is raised.
\item \textit{body~2}: the body that raises the tide.
\item \textit{tidal equilibrium}: a double synchronous state in which
  two bodies have a circular orbit with synchronized, aligned spins
  and no further tidal evolution occurs.
\item \textit{tidal potential}: the part of the gravitational
  potential due to body~2 that deforms body~1.
\item \textit{tidal component}: a single component of the tidal
  potential when it is decomposed into spherical harmonic functions
  and Fourier-analysed in time.
\item \textit{obliquity}: the angle between the rotation axis of
  body~1 and the orbit normal, also known as the spin--orbit
  misalignment.
\item \textit{inertial frame}: a non-rotating frame of reference.
\item \textit{fluid frame}: a frame of reference rotating with the
  spin angular velocity of body~1.
\item \textit{potential Love number}: a complex-valued, dimensionless,
  frequency-dependent response function that relates the
  self-gravitational perturbation of a body to the applied tidal
  potential.
\item \textit{tidal torque}: the rate of transfer of (axial) angular
  momentum from the orbit to the spin of body~1.
\item \textit{tidal power}: the rate of transfer of energy from the
  orbit to body~1.
\item \textit{(modified) tidal quality factor; phase lag; time lag}:
  alternative parametrizations of the tidal response functions, not to
  be taken literally.
\item \textit{f~mode}: a fundamental oscillation mode relying on the
  surface gravity of a body.
\item \textit{p~mode}: an acoustic mode relying on compressibility.
\item \textit{g~mode}: a global internal gravity mode relying on
  stable stratification.
\item \textit{internal wave}: a low-frequency, quasi-incompressible
  wave restored by buoyancy and/or Coriolis forces.
\item \textit{(internal) gravity wave}: an internal wave relying on
  buoyancy forces due to stable stratification.
\item \textit{inertial wave}: an internal wave relying on the Coriolis
  force due to rotation.
\item \textit{equilibrium tide}: a hydrostatic approximation to the
  perturbations of density and pressure, accompanied by a displacement
  and velocity that may not be uniquely defined.
\item \textit{dynamical tide}: a tidal response that satisfies the
  time-dependent equations of fluid dynamics, and usually includes
  internal waves.
\end{itemize}

\bigskip

\centerline{LITERATURE CITED}

\medskip

{\footnotesize

Adams ER, L\'opez-Morales M, Elliot JL, et al.\ 2011. \textit{Ap.\ J.} 741:102\\
Aerts C, Christensen-Dalsgaard J, Kurtz DW. 2010. Asteroseismology. Dordrecht: Springer\\
Albrecht S, Reffert S, Snellen I, Quirrenbach A, Mitchell DS. 2007. \textit{Astron.\ Astrophys.} 474:565--573\\
Albrecht S, Winn JN, Johnson JA, et al.\ 2012. \textit{Ap.\ J.} 757:18\\
Alexander ME. 1973. \textit{Astrophys.\ Space Sci.} 23:459--510\\
Arras P, Socrates A. 2010. \textit{Ap.\ J.} 714:1--12\\
Barker AJ. 2011. \textit{MNRAS} 414:1365--1378\\
Barker AJ, Lithwick Y. 2013. \textit{MNRAS} 435:3614--3626\\
Barker AJ, Lithwick Y. 2014. \textit{MNRAS} 447:305--315\\
Barker AJ, Ogilvie GI. 2010. \textit{MNRAS} 404:1849--1868\\
Baruteau C, Rieutord M. 2013. \textit{J.\ Fluid Mech.} 719:47--81\\
Bildsten L, Cutler C. 1992. \textit{Ap.\ J.} 400:175--180\\
Bodenheimer P, Lin DNC, Mardling RA. 2001. \textit{Ap.\ J.} 548:466--472\\
Bouchy F, Deleuil M, Guillot T. 2011. \textit{Astron.\ Astrophys.} 525:A68\\
Brown DJA, Collier Cameron A, Hall C, Hebb L, Smalley B. 2011. \textit{MNRAS} 415:605--618\\
Bryan GH. 1889. \textit{Phil.\ Trans.\ Roy.\ Soc.\ London A} 180:187--219\\
Burkart J, Quataert E, Arras P, Weinberg NN. 2012. \textit{MNRAS} 421:983--1006\\
Cartan E. 1922. \textit{Bull.\ Sci.\ Math.\ France} 46:317--352\\
Cartwright DE. 1999. Tides: A Scientific History.  Cambridge: Cambridge University Press\\
C\'ebron D, Le Bars M, Le Gal P, Moutou C, Leconte J, Sauret A. 2013. \textit{Icarus}, 226:1642--1653\\
C\'ebron D, Le Bars M, Leontini J, Maubert P, Le Gal P. 2010. \textit{Physics of the Earth and Planetary Interiors} 182:119--128\\
Charnoz S, Crida A, Castillo-Rogez JC, et al. 2011. \textit{Icarus} 216:535--550\\
Charnoz S, Salmon J, Crida A. 2010. \textit{Nature} 465:752--754\\
Claret A, Cunha NCS. 1997. \textit{Astron.\ Astrophys.} 318:187--197\\
Comeau S, Poisson E. 2009. \textit{Phys.\ Rev.\ D} 80:087501\\
Correia ACM, Laskar J, N\'eron de Surgy O. 2003. \textit{Icarus} 163:1--23\\
Counselman CC. 1973. \textit{Ap.\ J.} 180:307--316\\
Cowling TG. 1941. \textit{MNRAS} 101:367--375\\
Crida A, Charnoz S. 2012. \textit{Science} 338:1196--1199\\
Darwin GH. 1880. \textit{Phil.\ Trans.\ Roy.\ Soc.\ London} 171:713--891\\
Deparis V, Legros H, Souchay J. 2013. In \textit{Tides in Astronomy and Astrophysics}, ed.\ J Souchay, S Mathis, T Tokieda, pp.\ 31--82. Heidelberg: Springer\\
Dermott SF. 1979. \textit{Icarus} 37:310--321\\
Dintrans B, Rieutord M, Valdettaro L. 1999. \textit{J.\ Fluid Mech.} 398:271--297\\
Eckart C. 1960. Hydrodynamics of Oceans and Atmospheres.  Oxford: Pergamon\\
Eggleton PP, Kiseleva LG, Hut P. 1998. \textit{Ap.\ J.} 499:853--870\\
Fabian AC, Pringle JE, Rees MJ. 1975. \textit{MNRAS} 172:15p--18p\\
Favier B, Barker AJ, Baruteau C, Ogilvie GI. 2014. \textit{MNRAS} 493:845--60\\
French M, Becker A, Lorenzen W, et al. 2012. \textit{Ap.\ J.\ Supp.} 202:5\\
Fuller J, Lai D. 2012a. \textit{MNRAS} 420:3126--3138\\
Fuller J, Lai D. 2012b. \textit{MNRAS} 421:426--445\\
Gavrilov SV, Zharkov VN. 1977. \textit{Icarus} 32:443--449\\
Giersz M. 1986. \textit{Acta Astron.} 36:181--209\\
Gold T, Soter S. 1969. \textit{Icarus} 11:356--366\\
Goldreich P. 1963. \textit{MNRAS} 126:257--268\\
Goldreich P. 1965. \textit{MNRAS} 130:159--181 \\
Goldreich P, Keeley DA. 1977. \textit{Ap.\ J.} 211:934--942\\
Goldreich P, Nicholson PD. 1977. \textit{Icarus} 30:301--304\\
Goldreich P, Nicholson PD. 1989. \textit{Ap.\ J.} 342:1079--1084\\
Goldreich P, Soter S. 1966. \textit{Icarus} 5:375--389\\
Goodman J, Dickson ES. 1998. \textit{Ap.\ J.} 507:938--944\\
Goodman J, Lackner C. 2009. \textit{Ap.\ J.} 696:2054--2067\\
Goodman J, Oh SP. 1997. \textit{Ap.\ J.} 486:403--412\\
Greenspan H. 1968. The Theory of Rotating Fluids.  Cambridge: Cambridge University Press\\
Gu P-G, Lin DNC, Bodenheimer PH. 2003. \textit{Ap.\ J.} 588:509--534\\
Guillochon J, Ramirez-Ruiz E, Lin D. 2011. \textit{Ap.\ J.} 732:74\\
Hansen BMS. 2010. \textit{Ap.\ J.} 723:285--299\\
Hansen BMS. 2012. \textit{Ap.\ J.} 757:6\\
Hartle JB. 1973. \textit{Phys.\ Rev.\ D} 8:1010--1024\\
Hebb L, Collier Cameron A, Triaud AHMJ, et al. 2010. \textit{Ap.\ J.} 708:224--231\\
Hellier C, Anderson DR, Collier Cameron A, et al. 2009. \textit{Nature} 460:1098--1100\\
Hough SS. 1897. \textit{Phil.\ Trans.\ Roy.\ Soc.\ London A} 189:201--257\\
Hough SS. 1898. \textit{Phil.\ Trans.\ Roy.\ Soc.\ London A} 191:139--185\\
Husnoo N, Pont F, Mazeh T, et al. 2012. \textit{MNRAS} 422:3151--3177\\
Hut P. 1980. \textit{Astron.\ Astrophys.} 92:167--170\\
Hut P. 1981. \textit{Astron.\ Astrophys.} 99:126--140\\
Ibgui L, Burrows A. 2009. \textit{Ap.\ J.} 700:1921--1932\\
Ioannou PJ, Lindzen RS. 1993. \textit{Ap.\ J.} 406:266--278\\
Ivanov PB, Papaloizou JCB. 2007. \textit{MNRAS} 376:682--704\\
Ivanov PB, Papaloizou JCB. 2010. \textit{MNRAS} 407:1609--1630\\
Jackson B, Barnes R, Greenberg R. 2009. \textit{Ap.\ J.} 698:1357--1366\\
Jackson B, Greenberg R, Barnes R. 2008. \textit{Ap.\ J.} 678:1396--1406\\
Jeffreys H. 1961. \textit{MNRAS} 122:339--343\\
Jouve L, Ogilvie GI. 2014. \textit{J.\ Fluid Mech.} 745:223--250\\
Kaula WM. 1961. \textit{Geophys.\ J.\ Int.} 5:104--133\\
Kerswell RR. 2002. \textit{Annu.\ Rev.\ Fluid Mech.} 34:83–113\\
Khaliullin KhF, Khaliullina AI. 2007. \textit{MNRAS} 382:356--366\\
Khaliullin KhF, Khaliullina AI. 2010. \textit{MNRAS} 401:257--274\\
Kipping D. 2013. \textit{MNRAS} 434:L51--L55\\
Kochanek CS. 1992. \textit{Ap.\ J.} 398:234--247\\
Kopal Z. 1978. Dynamics of Close Binary Systems. Dordrecht: Reidel\\
Kosovichev AG, Novikov ID. 1992. \textit{MNRAS} 258:715--724\\
Kramm U, Nettelmann N, Fortney JJ, Neuha\"user R, Redmer R. 2012. \textit{Astron.\ Astrophys.} 538:A146\\
Kumar P, Goodman J. 1996. \textit{Ap.\ J.} 466:946--956\\
Lai D. 1997. \textit{Ap.\ J.} 490:847--862\\
Lai D. 2012. \textit{MNRAS} 423:486--492\\
Lainey V, Arlot J-E, Karatekin \"O, van Hoolst T. 2009. \textit{Nature} 459:957--959\\
Lainey V, Karatekin \"O, Desmars J, et al. 2012. \textit{Ap.\ J.} 752:14\\
Langer N. 2009. \textit{Astron.\ Astrophys.} 500:133--134\\
Laplace PS. 1775. \textit{M\'em.\ Acad.\ Roy.\ Sci.} 88:75--182\\
Le Bars M, Lacaze L, Le Diz\`es S, Le Gal P, Rieutord R. 2010. \textit{Physics of the Earth and Planetary Interiors} 178:48--55\\
Leconte J, Chabrier G, Baraffe I, Levrard B. 2010. \textit{Astron.\ Astrophys.} 516:A64\\
Lee HM, Ostriker JP. 1986. \textit{Ap.\ J.} 310:176--188\\
Leinhardt ZM, Ogilvie GI, Latter HN, Kokubo E. 2012. \textit{MNRAS} 424:1419--1431\\
Levrard B, Winisdoerffer C, Chabrier G. 2009. \textit{Ap.\ J.\ Lett.} 692:L9--L13\\
Lubow SH, Tout CA, Livio M. 1997. \textit{Ap.\ J.} 484:866--870\\
Maas LRM, Lam F-PA. 1995. \textit{J.\ Fluid Mech.} 300:1--41\\
McMahon JM, Morales MA, Pierleoni C, Ceperley DM. 2012. \textit{Rev.\ Mod.\ Phys.} 84:1607--1653\\
McMillan SLW, McDermott PN, Taam RE. 1987. \textit{Ap.\ J.} 318:261--277\\
Marcy GW, Butler RP, Williams E, et al. 1997. \textit{Ap.\ J.} 481:926--935\\
Mardling RA. 1995a. \textit{Ap.\ J.} 450:722--731\\
Mardling RA. 1995b. \textit{Ap.\ J.} 450:732--747\\
Mathieu RD. 2005. In \textit{Tidal Evolution and Oscillations in Binary Stars}, ed.\ A Claret, A Gim\'enez, J-P Zahn, pp.\ 26--38. San Francisco: Astronomical Society of the Pacific\\
Matsumura S, Peale SJ, Rasio FA. 2010. \textit{Ap.\ J.} 725:1995--2016\\
Meibom S, Mathieu RD. 2005. \textit{Ap.\ J.} 620:970--983\\
Meibom S, Mathieu RD, Stassun KG. 2006. \textit{Ap.\ J.} 653:621--635\\
Meyer J, Wisdom J. 2007. \textit{Icarus} 188:535--539\\
Mied RP. 1976. \textit{J.\ Fluid Mech.} 78:763--784\\
Ogilvie GI. 2005. \textit{J.\ Fluid Mech.} 543:19--44\\
Ogilvie GI. 2009. \textit{MNRAS} 396:794--806\\
Ogilvie GI. 2013. \textit{MNRAS} 429:613--632\\
Ogilvie GI, Lesur G. 2012. \textit{MNRAS} 422:1975--1987\\
Ogilvie GI, Lin DNC. 2004. \textit{Ap.\ J.} 610:477--509\\
Ogilvie GI, Lin DNC. 2007. \textit{Ap.\ J.} 661:1180--1191\\
O'Leary RM, Burkart J. 2014. \textit{MNRAS} 440:3036--3050\\
Papaloizou JCB, Ivanov PB. 2005. \textit{MNRAS} 364:L66--L70\\
Papaloizou JCB, Ivanov PB. 2010. \textit{MNRAS} 407:1631--1656\\
Papaloizou JCB, Pringle JE. 1978. \textit{MNRAS} 182:423--442\\
Papaloizou JCB, Savonije GJ. 1997. \textit{MNRAS} 291:651--657\\
Peale SJ. 1999. \textit{Annu.\ Rev.\ Astron.\ Astrophys.} 37:533--602\\
Penev K, Barranco J, Sasselov D, 2009a. \textit{Ap.\ J.} 705:285--297\\
Penev K, Sasselov D, Robinson F, Demarque P. 2007. \textit{Ap.\ J.} 655:1166--1171\\
Penev K, Sasselov D, Robinson F, Demarque P, 2009b. \textit{Ap.\ J.} 704:930--936\\
Poincar\'e H. 1885. \textit{Acta Math.} 7:259--380\\
Poisson E. 2009. \textit{Phys.\ Rev.\ D} 80:064029\\
Polfliet R, Smeyers P. 1990. \textit{Astron.\ Astrophys.} 237:110--124\\
Pont F. 2009. \textit{MNRAS} 396:1789--1796\\
Press WH, Teukolsky SA. 1977. \textit{Ap.\ J.} 213:183--192\\
Press WH, Wiita PJ, Smarr LL. 1975. \textit{Ap.\ J.\ Lett.} 202:L135--L137\\
Rappaport S, Sanchis-Ojeda R, Rogers LA, Levine A, Winn JN. 2013. \textit{Ap.\ J.\ Lett.} 773:L15\\
Rasio FA, Tout CA, Lubow SH, Livio M. 1996. \textit{Ap.\ J.} 470:1187--1191\\
Remus F, Mathis S, Zahn J-P, Lainey V. 2012. \textit{Astron.\ Astrophys.} 541:A165\\
Rieutord M. 1992. \textit{Astron.\ Astrophys.} 259:581--584\\
Rieutord M. 2004. In \textit{Stellar Rotation}, ed.\ A Maeder, P Eenens, pp.\ 394--403. San Francisco: Astronomical Society of the Pacific\\
Rieutord M, Georgeot B, Valdettaro L. 2001. \textit{J.\ Fluid Mech.} 435:103--144\\
Rieutord M, Valdettaro L. 2010. \textit{J.\ Fluid Mech.} 643:363--394\\
Rieutord M, Zahn J-P. 1997. \textit{Ap.\ J.} 474:760--767\\
Rogers TM, Lin DNC. 2013. \textit{Ap.\ J.\ Lett.} 769:L10\\
Sanchis-Ojeda R, Rappaport S, Winn JN, et al. 2013. \textit{Ap.\ J.} 774:54\\
Savonije GJ, Papaloizou JCB. 1983. \textit{MNRAS} 203:581--593\\
Savonije GJ, Papaloizou JCB. 1984. \textit{MNRAS} 207:685--704\\
Savonije GJ, Papaloizou JCB. 1997. \textit{MNRAS} 291:633--650\\
Savonije GJ, Papaloizou JCB, Alberts F. 1995. \textit{MNRAS} 277:471--496\\
Savonije GJ, Witte MG. 2002. \textit{Astron.\ Astrophys.} 386:211--221\\
Schlaufman KC, Lin DNC, Ida S. 2010. \textit{Ap.\ J.\ Lett.} 724:L53--L58\\
Schlaufman KC, Winn JN. 2013. \textit{Ap.\ J.} 772:143\\
Siverd RJ, Beatty TG, Pepper J, et al.\ 2012. \textit{Ap.\ J.} 761:123\\
Socrates A. 2013. arXiv 1304.4121\\
Sridhar S, Tremaine S. 1992. \textit{Icarus} 95:86--99\\
Stevenson DJ. 1983. \textit{J.\ Geophys.\ Res.} 88:2445--2455\\
Storch NI, Lai D. 2014. \textit{MNRAS} 438:1526--1534\\
Tassoul J-L. 1987. \textit{Ap.\ J.} 322:856--861\\
Terquem C, Papaloizou JCB, Nelson RP, Lin DNC. 1998. \textit{Ap.\ J.} 502:788--801\\
Thompson SE, Everett M, Mullally F, et al. 2012. \textit{Ap.\ J.} 753:86\\
Thomson W. 1863. \textit{Phil.\ Trans.\ Roy.\ Soc.\ London} 153:583--616\\
Thomson W. 1880. \textit{Phil.\ Mag.} 10:155--168\\
Thorne KS, Price RH, MacDonald DA. 1986. Black Holes: The Membrane Paradigm. New Haven: Yale University Press\\
Venumadhav T, Zimmerman A, Hirata CM. 2014. \textit{Ap.\ J.} 781:23\\
Verbunt F, Phinney ES. 1995. \textit{Astron.\ Astrophys.} 296:709--721\\
Watson CA, Marsh TR. 2010. \textit{MNRAS} 405:2037--2043\\
Weinberg NN, Arras P, Burkart J. 2013. \textit{Ap.\ J.} 769:121\\
Weinberg NN, Arras P, Quataert E, Burkart J. 2012. \textit{Ap.\ J.} 751:136\\
Welsh WF, Orosz JA, Aerts C, et al. 2011. \textit{Ap.\ J.\ Supp.} 197:4\\
Welsh WF, Orosz JA, Seager S, et al. 2010. \textit{Ap.\ J.\ Lett.} 713:L145--L149\\
Winn JN, Fabrycky D, Albrecht S, Johnson JA. 2010. \textit{Ap.\ J.\ Lett.} 718:L145--L149\\
Witte MG, Savonije GJ. 1999a. \textit{Astron.\ Astrophys.} 341:842--852\\
Witte MG, Savonije GJ. 1999b. \textit{Astron.\ Astrophys.} 350:129--147\\
Witte MG, Savonije GJ. 2001. \textit{Astron.\ Astrophys.} 366:840--857\\
Witte MG, Savonije GJ. 2002. \textit{Astron.\ Astrophys.} 386:222--236\\
Wu Y. 2005a. \textit{Ap.\ J.} 635:674--687\\
Wu Y. 2005b. \textit{Ap.\ J.} 635:688--710\\
Yoder CF, Peale SJ. 1981. \textit{Icarus} 47:1--35\\
Zahn J-P. 1966a. \textit{Ann.\ Astrophys.} 29:313--330\\
Zahn J-P. 1966b. \textit{Ann.\ Astrophys.} 29:489--506\\
Zahn J-P. 1970. \textit{Astron.\ Astrophys.} 4:452--461\\
Zahn J-P. 1975. \textit{Astron.\ Astrophys.} 41:329--344\\
Zahn J-P. 1977. \textit{Astron.\ Astrophys.} 57:383--394\\
Zahn J-P. 1989. \textit{Astron.\ Astrophys.} 220:112--116\\
Zahn J-P, Bouchet L. 1989. \textit{Astron.\ Astrophys.} 223:112--118\\

}
\end{document}